\documentclass[%
 reprint,
nofootinbib,
 amsmath,amssymb,
 aps,
]{revtex4-2}

\usepackage{graphicx}
\graphicspath{{figures/}}
\usepackage{dcolumn}
\usepackage{bm}
\usepackage{hyperref}

\usepackage{xcolor}
\usepackage{cleveref}

\newcommand{\avg}[1]{\langle #1 \rangle}
\newcommand{\gm}{\gamma}
\newcommand{\wt}{\widetilde}

\begin{document}


\title{Survival probability and size of lineages in antibody affinity maturation}

\author{Marco Molari}
\author{R\'{e}mi Monasson}
\author{Simona Cocco}
 \thanks{Email: simona.cocco@phys.ens.fr}
\affiliation{Laboratoire de Physique de l'\'{E}cole Normale Sup\'{e}rieure, ENS, PSL University, CNRS UMR8023, Sorbonne Universit\'{e}, Universit\'{e} de Paris, 24 rue Lhomond, 75005 Paris, France.}%

\date{\today}

\begin{abstract}
Affinity Maturation (AM) is the process through which the immune system is able to develop potent antibodies against new pathogens it encounters, and is at the base of the efficacy of vaccines. At its core AM is analogous to a Darwinian evolutionary process, where B-cells mutate and are selected on the base of their affinity for an Antigen (Ag), and Ag availability tunes the selective pressure. In cases when this selective pressure is high the number of B-cells might quickly decrease and the population might risk extinction in what is known as a \textit{population bottleneck}. Here we study the probability for a B-cell lineage to survive this bottleneck scenario as a function of the progenitor affinity for the Ag. Using recursive relations and probability generating functions we derive expressions for the average extinction time and progeny size for lineages that go extinct. We then extend our results to the full population, both in the absence and presence of competition for T-cell help, and quantify the population survival probability as a function of Ag concentration and initial population size. Our study suggests the population bottleneck phenomenology might represent a limit case in the space of biologically plausible maturation scenarios, whose characterization could help guide the process of vaccine development.
\end{abstract}

\maketitle

\section{Introduction}

\begin{figure}[b]
  \centering
  \includegraphics[width=0.5\textwidth]{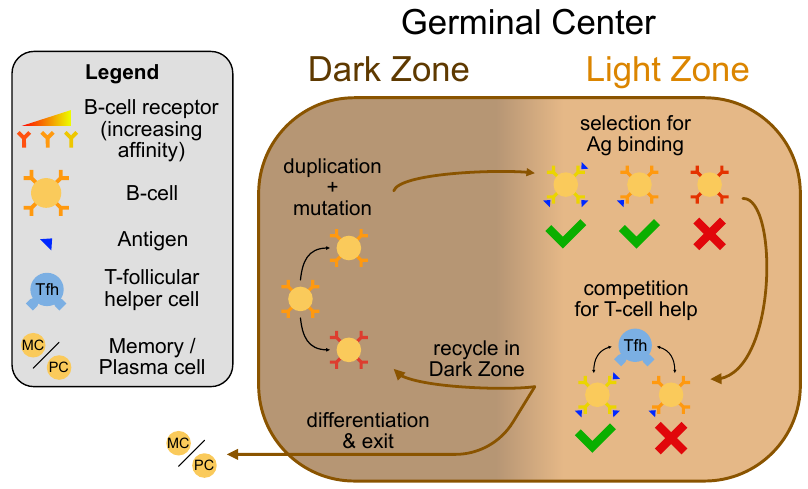}
  \caption{
  schematic depiction of the germinal center reaction. Inside of a Germinal Center (GC) B-cells undergo iterative cycles of duplication, mutation and selection. Cell duplication and mutation occurs in the GC dark zone, while selection takes place in the light zone. Selection is divided in two steps: cells must first bind the antigen with their B-cell receptors, and then compete to receive a survival signal from T-follicular helper cells. Failure in any of these steps results in apoptosis. Successfully selected cells have some probability of differentiating into Memory or Plasma Cells (MCs/PCs), and exit the GC. Cells that do not differentiate recycle back in the dark zone to start a new maturation cycle.
  }
  \label{fig:gc_scheme}
\end{figure}

Affinity Maturation (AM) is a biological process through which our immune system generates potent Antibodies (Ab) against newly-encountered pathogens. AM is also at the base of the efficacy of vaccination, in which this process is artificially elicited through the administration of a dose of Antigen (Ag). The biological mechanisms that govern AM are many and complex, and are the object of many excellent reviews \cite{victora2012germinal, de2015dynamics, bannard2017germinal, mesin2016germinal, eisen2014affinity, victora2014clonal, shlomchik2019linking, victora2018primary}. Simply speaking, AM works by subjecting a population of B-lymphocytes (or B-cells) to iterative cycles of mutations and selection for Ag binding, which generate a Darwinian evolutionary process that progressively increases their affinity for the Ag. This process is schematically depicted in \cref{fig:gc_scheme}. Maturation takes place in Germinal Centers (GCs), microanatomical structures that appear inside of secondary lymphoid organs. They are divided in two areas: the GC Dark Zone (DZ) in which cells divide and mutate,\footnote{In the DZ B-cells express high levels of \textit{Activation-Induced cytidine Deaminase}, an enzyme that increases the natural rate of DNA mutations up to $10^{-3}$ per base-pair per generation \cite{kleinstein2003estimating,mckean1984generation,berek1987mutation}.} and the Light Zone (LZ) in which they undergo selection. Cells iteratively migrate between these two compartments. Selection in the LZ is completed in two steps. In the first step cells try to bind the Ag, exposed on the surface of Follicular Dendritic Cells (FDCs). In the second step they compete to receive a survival signal from T-follicular helper (Tfh) cells, in the absence of which they undergo apoptosis. Tfh cells are able to probe the amount of Ag captured by B-cells, preferentially delivering this survival signal to the cells that were most successful in capturing Ag. Cells that receive the signal either migrate back to the DZ for additional rounds of mutation and selection, or they can differentiate in Plasma or Memory Cells (PCs/MCs). The former are responsible for the production of Abs to fight the infection, while the latter confer long-lasting protection by remaining quiescent until the same Ag is encountered again, in which case they reactivate and produce Abs or enter GCs for further maturation.

In spite of the many recent experimental advancements in the study of AM, several open questions still remain to be answered, which have important implications in vaccine design.
For example understanding the role of Ag availability in controlling maturation might lead to optimization of Ag dosage in vaccines \cite{rhodes2019dose, molari2020elife, kang2015affinity}, or understanding the effect of B-cell precursor frequency and affinity might help improving immunogen design \cite{abbott2018precursor, abbott2020factors}.
Given the complexity of this process, computational models represent an invaluable tool to guide our understanding of AM \cite{chakraborty2017perspective, buchauer2019calculating}. In this paper we introduce a stochastic model of AM to study the survival probability of B-cell lineages in GCs. Experimental analysis of vaccine-responsive lineages shows signatures of selection in their reconstructed phylogenies \cite{horns2019signatures}. This selection pressure, which is partially controlled by Ag availability \cite{molari2020elife}, is important to push lineages towards maturation, but at the same time an excessive pressure might be deleterious. Indeed, several maturation models present a phenomenology termed \textit{population bottleneck} \cite{zhang2010optimality, wang2015manipulating, wang2017optimal}, in which strong selection pressure causes a decrease in GC population size, potentially leading to extinction. As a consequence of this trade-off optimal maturation is achieved at intermediate levels of selection pressure. While the bottleneck phenomenology has been studied through numerical simulations so far, we think that it is not incompatible with experimental evidence \cite{firl2018capturing}, and  might represent a limiting regime for AM as later discussed in \cref{subs:bottleneck_bio}.\\

The study of the survival probability of a population through the maturation bottleneck we report below is both analytical and numerical.  We start by considering the dependence of a lineage survival probability on the progenitor affinity. Through the use of recursive relations and probability generating functions we are able to evaluate this probability, and also quantify extinction time and progeny size for lineages that go extinct. We then extend our approach to analyze the extinction probability for the full B-cell population, and its dependence on Ag concentration and initial population size. Last of all, we discuss the biological relevance of the bottleneck phenomenology, and how quantifying lineage survival probability might help vaccine design.

\section{Model for stochastic maturation}

Our model for stochastic maturation is inspired by previous works \cite{wang2015manipulating,molari2020elife}. The model is simple enough to be analytically tractable, while retaining the main aspects of the bottleneck phenomenology. 

\subsection{Steps in affinity maturation}

We consider the evolution of a population of B-cells inside a GC. Through repeated cycles of mutation and selection the population increases its average affinity for the Ag over time. In our model each cell in the population is solely characterized by its affinity for the Ag, measured in terms of binding energy $\epsilon$ and expressed in units of $k_B T$.

\begin{figure*}[t]
  \centering
  \includegraphics[width=0.65\textwidth]{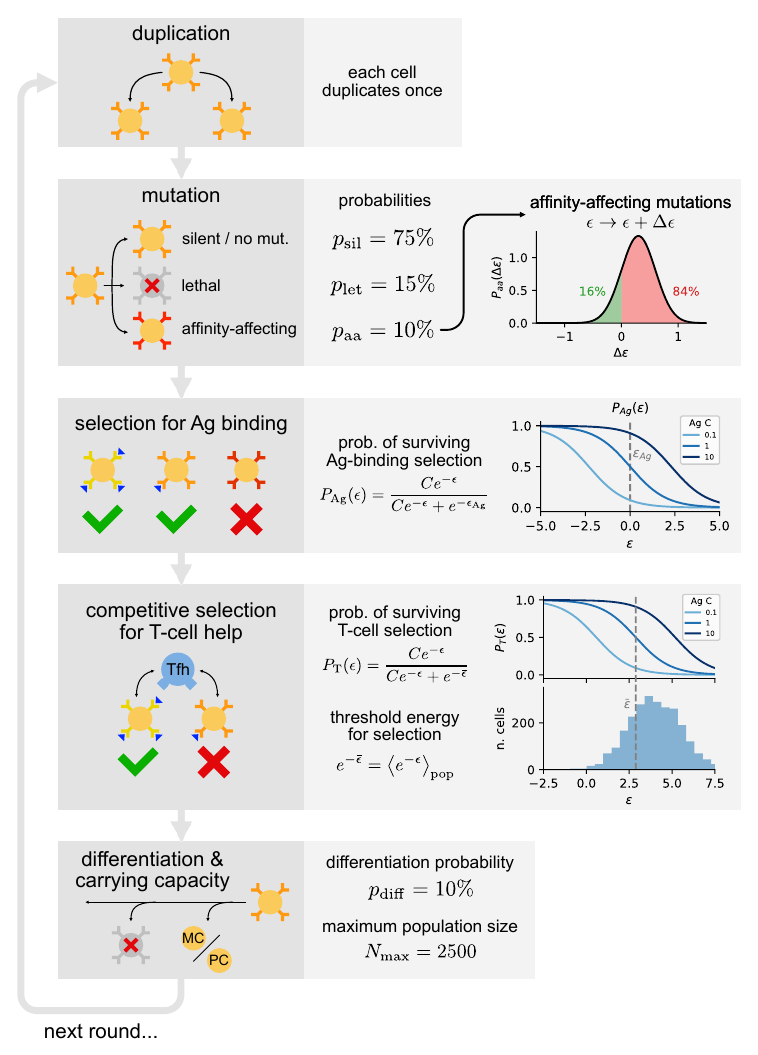}
  \caption{schematic description of the processes that make up a simulated evolution round in our model. At the beginning of the round cells duplicate once. Each cell can then independently develop a mutation. Cells that undergo a lethal mutation ($p_\mathrm{let} = 15\%$) are removed from the population, while cells that develop an affinity-affecting mutation ($p_\mathrm{aa} = 10\%$) receive an additive change $\Delta\epsilon$ to their binding energy, extracted from the displayed Gaussian distribution. Notice that most of the mutations have a deleterious effect on affinity. Cells undergo then selection for antigen binding and compete to receive T-cell help. For each selection step we display the functional behavior of the probability of surviving as a function of the progenitor affinity $\epsilon$ and antigen concentration $C$. In selection for T-cell help competition is obtained by making the threshold energy $\bar\epsilon$ depend on the current affinity distribution of the population, as displayed. Cells that are able to survive selection have a probability $p_\mathrm{diff} = 10\%$ of differentiating into memory or plasma cells, and exiting the cycle. Last of all, if the population size exceeds the threshold value $N_{\max} = 2500$ cells in excess are randomly removed. The remaining cells will then begin a new evolution round.}
  \label{fig:model_descr}
\end{figure*}

The simulation starts when the GC is mature (roughly 1 week after Ag injection \cite{de2015dynamics}). The initial population is composed of $N_i$ cells whose binding energy is independently extracted from a Gaussian distribution of naive responders, with mean $\mu_i$ and standard deviation $\sigma_i$. Cells undergo iterative rounds of duplication, mutation and selection. These steps are schematized in \cref{fig:model_descr}.


At the beginning of the round cells duplicate once in the GC DZ. Each daughter cell can then independently either:
\begin{itemize}
\item undergo an affinity-affecting mutation with probability $p_\mathrm{aa}$, which causes its binding energy to change by some amount $\Delta\epsilon$.  We assume that $\Delta \epsilon$ is a random variable, extracted from a Gaussian distribution with mean $\mu_\mathrm{M}$ and standard deviation $\sigma_\mathrm{M}$ (see \cref{fig:model_descr} ``mutation'');
\item not mutate or develop silent mutations, with probability $p_\mathrm{sil}$. In both cases its affinity is unchanged;
\item be hit by a lethal mutation with probability $p_\mathrm{let}$, in which case the cell it is removed from the population.
\end{itemize}
As a result, the total distribution of the changes $\Delta\epsilon$ is therefore summarized by the kernel
\begin{equation}\label{eq:mut_ker}
	K (\Delta \epsilon) =  \frac {p_\mathrm{aa}}{\sqrt{2\pi \sigma_M^2}} \;\exp\left( -\frac {(\Delta \epsilon-\mu_M)^2}{2\,\sigma_M^2}\right) + p_\mathrm{sil} \; \delta (\Delta \epsilon)
\end{equation}
where $\delta (\Delta \epsilon)$ is Dirac delta distribution. Notice that, due to lethal mutations, the integral of the kernel $K$ is not normalized to unity but to $p_\mathrm{aa}+p_\mathrm{sil}=1 - p_\mathrm{let}$. Parameters are chosen such that only a small fraction of the mutations is beneficial, i.e. decreases the binding energy (cf. \cref{app:parameters}).


After duplication and mutation cells migrate to the LZ where they try to bind the Ag exposed on the surface of FDCs. Failure to do so results in cell death, and only cells that are able to bind the Ag with sufficient affinity survive this step of selection. Similarly to \cite{wang2015manipulating,molari2020elife} we consider the survival probability for a cell with binding energy $\epsilon$ to be given by the following Langmuir isotherm:
\begin{equation} \label{eq:psurv_ag}
    P_\text{Ag}(\epsilon) = \frac{C e^{-\epsilon}}{C e^{-\epsilon} + e^{-\epsilon_\text{Ag}}}\ ,
\end{equation}
where $\epsilon_\text{Ag}$ is a threshold binding energy and $C$ represents the dimensionless concentration of Ag available for cells to bind. This concentration controls the strength of selection, making successful binding more likely when more Ag is available to bind. In practice it acts by imparting a shift of magnitude $\log C$ to the energy threshold. The functional dependence of the selection probability on $\epsilon$ and $C$ is displayed in \cref{fig:model_descr} (``selection for Ag binding'').

In a second selection step cells compete to receive a survival signal from T-follicular helper cells, with the signal being preferentially delivered to cells that bind more Ag. The survival probability for a cell with binding energy $\epsilon$ is:
\begin{equation} \label{eq:psurv_t}
    P_\text{T}(\epsilon, \bar\epsilon) = \frac{C e^{-\epsilon}}{C e^{-\epsilon} + e^{-\bar\epsilon}}, \qquad \mathrm{with} \; e^{-\bar\epsilon} = \avg{e^{-\epsilon}}_\mathrm{pop}
\end{equation}
Where the term $\avg{e^{-\epsilon}}_\mathrm{pop}$ represents the average of this quantity over the population and encodes for competition (see \cref{fig:model_descr} ``competitive selection for T-cell help''). The surviving cells can then differentiate into plasma or memory cells with total probability $p_\mathrm{diff}$. We do not keep track of these differentiated cells in the simulation.

After this step if the population size exceeds the maximum carrying capacity $N_{\max}$ cells are randomly removed until this threshold is met. The surviving cells start then the next round of evolution. The values of the model parameters are reported in \cref{tab:model_par}, and discussed in \cref{app:parameters}.

\subsection{Population bottleneck and lineage survival}

\begin{figure*}[t]
  \centering
  \includegraphics[width=0.9\textwidth]{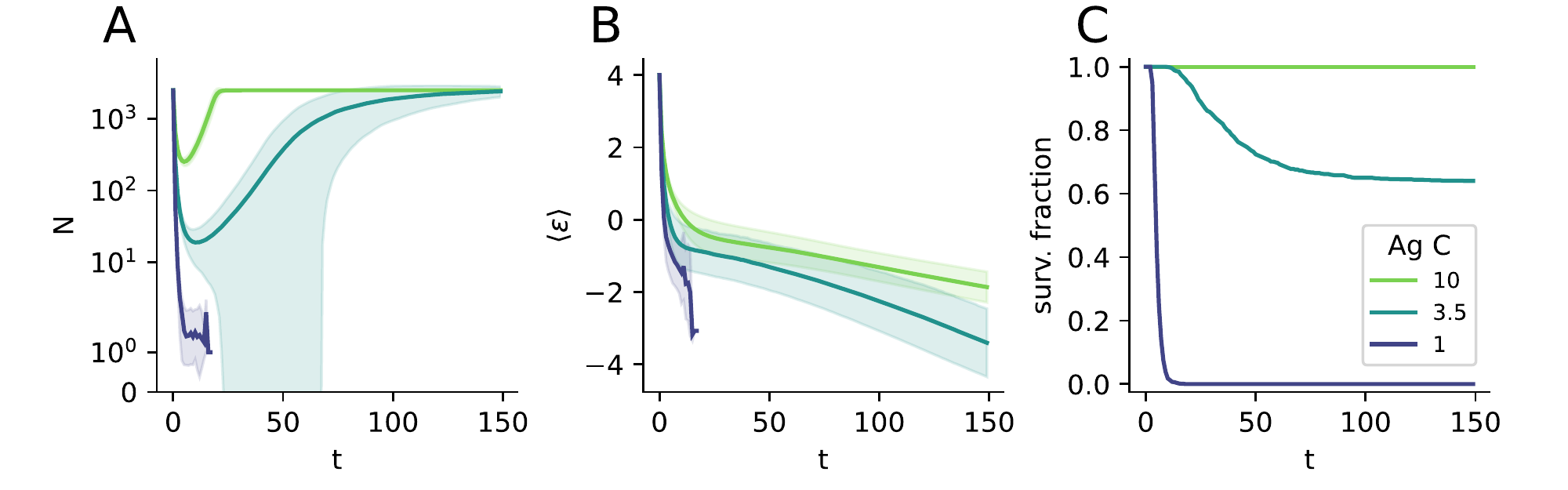}
  \caption{Average evolution of 1000 different stochastic simulations of the model at three different levels of Ag concentration $C = 1, 3.5, 10$, color-coded according to the legend on the right. \textbf{A}: population size $N$ as a function of evolution round. Shaded area covers one standard deviation for surviving simulations. The minimum population size on the bottleneck depends strongly on Ag concentration \textbf{B}: same as panel A but for the average population binding energy $\avg{\epsilon}$. Notice how for surviving populations the maturation speed depends on Ag concentration. \textbf{C}: Fraction of surviving simulations as a function of time. At low concentration the bottleneck drives all simulations to extinction, while at high concentration the population survives with high probability.}
  \label{fig:evo_example}
\end{figure*}

\begin{figure*}[t]
  \centering
  \includegraphics[width=0.9\textwidth]{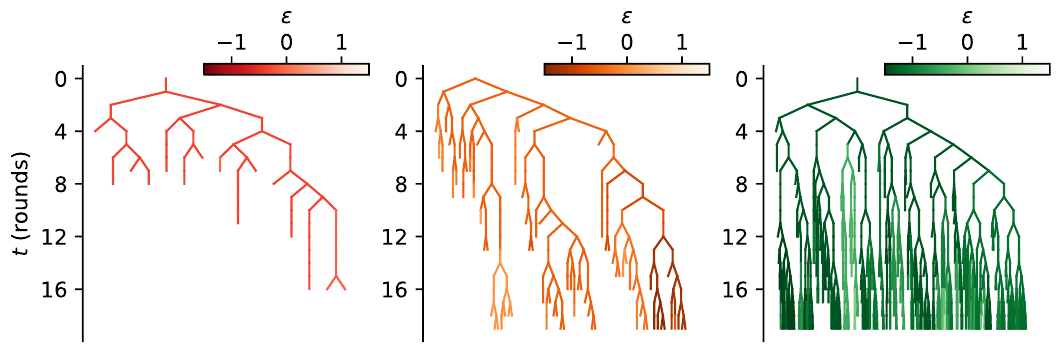}
  \caption{Examples of stochastic lineage evolution through a population bottleneck. We perform a single simulation of our model at Ag concentration $C=5$ and consider three different progenitors with different initial affinities (red $\epsilon_i \sim -0.3$, orange $\epsilon_i \sim -0.45$ and green $\epsilon_i \sim -1.3$). We represent their progeny evolution in the form of a tree with each cell corresponding to a node, and encoding affinity in the branch color. The lineage of the red progenitor quickly goes extinct, while the lineage of the orange progenitor survives the bottleneck but only with few individuals. The green progenitor lineage conversely survives the population bottleneck and undergoes great expansion. Notice how fate correlates with the initial progenitor affinity.}
  \label{fig:phylogenies_example}
\end{figure*}

Similarly to other AM models \cite{wang2015manipulating,zhang2010optimality}, for standard parameter values the population initially undergoes a bottleneck state. This is caused by the strong selection pressure initially imposed by Ag-binding selection, which later relaxes if the average population energy reaches values $\avg{\epsilon}_\text{pop} \sim \epsilon_\text{Ag}$. By controlling the selection pressure (cf. \cref{eq:psurv_ag,eq:psurv_t}) Ag concentration also impacts the population survival probability. 

As an illustration we report in \cref{fig:evo_example} the average evolution of 1000 stochastic simulations for three different values of the concentration $C$. For all three values the population size initially decreases under the combined effect of the two selection steps (\cref{fig:evo_example}A). This decrease lasts for few evolution rounds, and is accompanied by a quick increase in average affinity (\cref{fig:evo_example}B). At this point surviving populations are composed of few high-affinity cells, on which the main acting selection force  is competitive selection in  \cref{eq:psurv_t}. If this selection pressure is not too strong then the population will later expand and mature. Through a mechanism analogous to the one studied in \cite{molari2020elife} Ag concentration then controls the maturation speed, as can be seen by comparing the speed of decrease in average binding energy of the population after the bottleneck in the cases $C = 10$ and $C=3.5$ in \cref{fig:evo_example}B. 

The fraction of surviving simulations as a function of time is shown in  \cref{fig:evo_example}C. At low concentration ($C=1$) the population goes quickly extinct in all simulations. For such small values of Ag concentration competitive selection alone is sufficiently strong to impede population growth. Intermediate concentration values ($C=3.5$), on the contrary, are sufficient to sustain population growth. In this case extinction can nevertheless occur close to the bottleneck state, when population size  gets transiently small, see \cref{fig:evo_example}A; if some cells are able to survive and pass this bottleneck, then the population again grows  to full size and continues maturation. Last of all, at high concentration ($C=10$), the bottleneck pressure is not sufficient to significantly endanger population survival, and all simulations are able to overcome the low-population state without going extinct, but maturation proceeds more slowly. 
We will study in detail in the next section the dependence of the survival probability of the population of cells on Ag concentration.

Survival and future expansion is also strongly dependent on the initial distribution of affinities. This effect can be readily observed on lineages originated from a single ancestor, with energy $\epsilon$. In \cref{fig:phylogenies_example} we display three examples of lineage evolution in the form of trees in which each node corresponds to a different cell. These lineages differ by the affinity of their progenitor at the root of the tree. The progeny of the lowest-affinity one (red, $\epsilon_i \sim -0.3$) goes extinct in few evolution rounds. In the one with intermediate affinity (orange, $\epsilon_i \sim -0.45$) only few individuals are able to survive the bottleneck. The high-affinity one (green, $\epsilon_i \sim -1.3$) instead expands and eventually takes over the population. To quantify the population survival probability we will first investigate how the survival of single lineages depends on the progenitor affinity.

\section{Probability of survival and distribution of extinction times}
\label{sec:surv_prob}

In this section we study the probability that a B-cell lineage descending from a single progenitor cell survives through a population bottleneck, in particular how the probability of survival depends on the affinity of the progenitor. We also determine the distribution of extinction times of the lineage. We then make use of these results to evaluate the survival probability for the full population.

\subsection{Case of one lineage}
\label{subs:ext_prob_and_time}

\subsubsection{Probability of survival}

\begin{figure*}
  \includegraphics[width=0.9\textwidth]{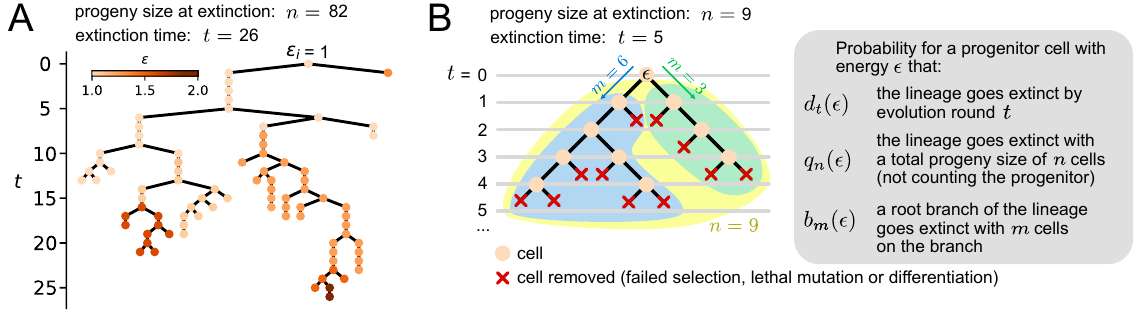}
  \caption{\textbf{A}. Example of lineage  issued from a progenitor with binding energy $\epsilon_i=1$ obtained from a stochastic simulation performed at Ag concentration $C=7$ in the approximation of only Ag-binding selection. Each node in the tree represents a cell, its binding energy $\epsilon$ encoded using the colorscale on top. In this example cells progressively accumulate deleterious mutations until after 26 evolution rounds the lineage eventually goes extinct.
  \textbf{B}. Schematic illustration of the quantities analyzed in our theory. On the left we depict a lineage evolution, stemming from a progenitor with binding energy $\epsilon$. The probability that such a lineage goes extinct by time $t$ is indicated with $d_t(\epsilon)$ (in this example $t=5$). The quantity $q_n(\epsilon)$ represents instead the probability that the lineage goes extinct counting a total of $n$ cells not including the progenitor ($n = 9$ in this example). Finally, $b_m(\epsilon)$ is the probability that one of the two sub-lineage stemming from the progenitor goes extinct counting $m$ cells (here $m=6$ for the left branch and $m=3$ for the right one).}
  \label{fig:explanation}
\end{figure*}

Let us consider a progenitor cell with binding energy $\epsilon$, present in the population at the beginning of evolution $t=0$. At each evolution round this cell will divide and its offspring will have some probability of being removed from the population, either due to selection, differentiation or lethal mutations. Since we are interested in studying the bottleneck phase, in which the population is not at its maximum size, we can neglect the enforcement of the carrying capacity constraint. In \cref{fig:explanation}A we report an example of lineage evolution for a progenitor with binding energy $\epsilon=1$. Color indicates the binding energy of each cell, according to the color-scale on top. In this example cells accumulate deleterious mutations until the lineage eventually goes extinct after $t=26$ evolution rounds.

We are interested in computing the probability $d_t(\epsilon)$ that \textit{all of the offspring} of a progenitor with binding energy $\epsilon$ will be extinct by evolution round $t$, see \cref{fig:explanation}B.
The expression for $t=1$ can easily be written as the probability that both daughter cells generated during the duplication phase will be removed by the end of the round. As stated above, this can occur either by lethal mutation, by failing selection or by differentiation. For each daughter cell this probability is more easily expressed as one minus the probability of not being removed:
\begin{equation}\label{eq:d1}
	d_1(\epsilon) = \left[1 - \int d\Delta\epsilon \, K(\Delta\epsilon) \, P_S(\epsilon + \Delta\epsilon) \, (1 - p_\mathrm{diff})\right]^2\ ,
\end{equation}
where the expression for $K(\Delta)$ is the one given in \cref{eq:mut_ker}, and $P_S(\epsilon)$ is the probability for a cell with binding energy $\epsilon$ of surviving selection. 
In the bottleneck state most of the selection pressure is generated by Ag-binding selection (i.e. $\bar\epsilon_t < \epsilon_\mathrm{Ag}$). As a first approximation we therefore neglect competitive selection for T-cell help, and consider simply $P_S(\epsilon) = P_\text{Ag}(\epsilon)$ (cf. \cref{eq:psurv_ag}). This introduces two important simplifications. First, the expression of $P_S$ does not depend on time.  Second, removing the competitive selection decouples the fate of all cells in the population.\\
The probabilities $d_t(\epsilon)$ for $t>1$ can be evaluated using recursive relations that express the probability of extinction in $t$ rounds as the probability for each daughter cell to either go extinct in one round, or to survive the first round but to have their respective offspring go extinct in the remaining $t - 1$ rounds:
\begin{multline}\label{eq:dt_recursive}
	d_t(\epsilon) = \bigg[ 1 - \int d\Delta\epsilon \, K(\Delta\epsilon) \, P_S(\epsilon + \Delta\epsilon) \, (1 - p_\mathrm{diff}) \\
	\times  \, \left(1 - d_{t-1}(\epsilon + \Delta\epsilon) \right)\bigg]^2
\end{multline}
In other words, the probability that all of the offspring goes extinct in $t$ rounds is the probability that each of the lineages stemming from the two daughter cells generated during the duplication phase of the first round are  removed before the end of round $t$. Since division is symmetric this probability must be the same for each daughter cell, and is the term inside the square brackets. This term is more easily expressed as one minus the probability that the lineage survives (term in the integral). In turn this can be decomposed as the probability that the daughter cell survives mutation (possibly with an energy change of entity $\Delta\epsilon$), selection and differentiation, multiplied by the probability that its offspring does not go extinct by the following $t-1$ rounds (term on the second line). A more extended explanation for the derivation of this equation and its numerical evaluation is provided in \cref{app:rec_explanation}.

\begin{figure*}
  \includegraphics[width=0.5\textwidth]{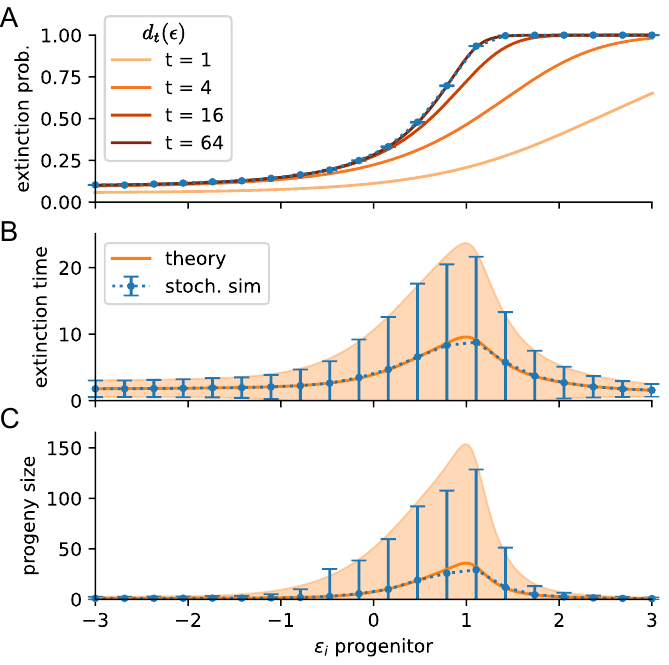}
  \caption{Comparison between stochastic simulations (blue) and theory (orange) for the probability of extinction (A), lineage extinction time (B) and average progeny size at extinction (C) as a function of the progenitor energy $\epsilon_i$ in absence of competitive selection. For each conditions we consider 5000 different stochastic simulations that terminate with extinction at Ag concentration $C=7$. \textbf{A}: stochastic extinction probability (blue dots, error bar indicate the standard error of the mean) evaluated as the fraction of simulations that terminate with extinction over the total number of simulations performed. This is compared to the value of $d_t(\epsilon)$ as described by our theory. 
  \textbf{B}: mean and standard deviation of extinction time (blue) over 5000 simulations terminating in extinction. This is compared to the theoretical prediction (orange) for the mean and standard deviation of this quantity, obtained using the time extinction probability $r_t(\epsilon)$. \textbf{C}: same as B but for the progeny size. In this case the theoretical predictions are obtained using the generating function theory.}
  \label{fig:tree_stats}
\end{figure*}

In \cref{fig:tree_stats}A we plot the behavior of $d_t(\epsilon)$ as a function of evolution round $t$ and progenitor binding energy $\epsilon$ (orange curves, color indicates extinction round $t$). As expected, the extinction probability is a monotonically increasing function of time and of energy, and reaches an asymptotic value $d_{\infty}(\epsilon)$ for large $t$. Our analytical result is in excellent agreement with simulations for the mean extinction probability (blue dots). The asymptotic probability $d_{\infty}(\epsilon)$ ranges between:
\begin{itemize}
    \item $d_{\infty}(\epsilon\to +\infty) =1$. This is easily understood, since high-energy (i.e. low-affinity) cells will not pass the selection step and their progeny will quickly go extinct.
    \item $d_{\infty}(\epsilon\to -\infty) =\min \{1, [1 - 1/\alpha]^2 \}$, with $\alpha = (1 - p_\mathrm{let}) (1-p_\mathrm{diff})$.
    The value of extinction probability $d_\infty$ for very high-affinity cells may be higher than 0 since lethal mutations and differentiation may still drive the lineage to extinction, especially during the first few evolution rounds when the offspring size is still small.
    The above expression for $d_{\infty}(\epsilon\to -\infty)$ can be obtained by searching a fixed point to \cref{eq:dt_recursive},   and considering that, for $\epsilon \to -\infty$, mutations do not sensibly change the survival probability and can  be neglected. The parameter $\alpha$ defined above is then the probability for a high-affinity cell to survive one round and not be removed by lethal mutations or through differentiation. Notice that, if $\alpha < 1/2$, we have $d_\infty (\epsilon \to +\infty) = 1$, as is to be expected when on average less than one individual in the offspring will survive. However this case is pathological: in this regime, irrespective of the progenitor energy, the population always goes extinct ($d_{\infty}(\epsilon) = 1$).
\end{itemize}

It is worthwhile to notice that the “infinite-time” extinction probability $d_\infty$ described in our theory represents the probability for a lineage to go extinct in the bottleneck phase, and does not reflect the probability of lineage fixation or extinction when the population is at maximum size. 
In \cref{eq:d1,eq:dt_recursive} we indeed neglected the carrying capacity constraint. This is justified when considering the bottleneck state, in which the population is not at its maximum size. However if some lineages are able to survive this state, then the population will eventually grow back to maximum size, and some cells will be randomly removed through enforcement of the carrying capacity. One should therefore extend the theory and add another term to correctly describe the lineage survival probability far away from the bottleneck.
\\

\subsubsection{Distribution of extinction times}

The probability that a lineage generated by a progenitor with energy $\epsilon$ goes extinct exactly at round $t$ can easily be expressed as 
\begin{equation}
\label{eq:rt}
r_t(\epsilon) = d_t(\epsilon) - d_{t-1}(\epsilon) \ .
\end{equation}
This allows us to evaluate the mean and variance for the extinction time probabilities (see \cref{fig:tree_stats}B) simply from the first two moments of the distribution:
\begin{equation}
	\avg{t}_{\epsilon} = \sum_{t=0}^{\infty} t \, r_t(\epsilon), \qquad 	\avg{t^2}_{\epsilon} = \sum_{t=0}^{\infty} t^2 \, r_t(\epsilon), \qquad
\end{equation}
In \cref{fig:tree_stats}B we compare, in the approximation of no competitive selection,  the average extinction time computed from simulations (blue, error bars indicate the standard deviation of extinction times for each progenitor affinity) with theoretical predictions  (orange, shaded area covers one standard deviations). We again find a very good match. The average extinction time shows a peak for intermediate affinities, which can be interpreted as follows. Low-affinity progenitors, i.e. having high binding energy have close-to-one probability of extinction, and very often go extinct in the first few rounds. High-affinity cells on the contrary have a small but non-zero probability of extinction, see value of extinction probability in \cref{fig:tree_stats}A. This is  mainly due to affinity-independent terms (differentiation and lethal mutation probabilities) which confer to the lineage a small chance of going extinct during the first few evolution rounds, when the progeny size is still small.
For affinities close to $\epsilon = 1$ we observe intermediate values for the probability of survival, and maximum value for average extinction time.  This behaviour can be better understood when mutations are turned off, in which case equations can be solved exactly, as shown below.

\subsubsection{Exactly solvable case of no mutation}

We hereafter consider the case of no affinity-affecting mutations, for which the mutation kernel \cref{eq:mut_ker} reads
\begin{equation} \label{eq:mut_ker_nomut}
    K(\Delta) = (1 - p_\mathrm{let}) \, \delta(\Delta)\ .
\end{equation}
In this case genealogies belong to the class of Galton-Watson trees \cite{watson1875probability}, and the asymptotic survival probability can be derived exactly. This probability is better expressed by considering the quantity
\begin{equation} \label{eq:gamma_nomut}
    \gamma(\epsilon) = (1 - p_\mathrm{let})  \, P_S(\epsilon) \, (1 - p_\mathrm{diff})
\end{equation}
which represents the probability for a daughter cell to remain in the GC and not be removed by either lethal mutations, selection or differentiation. The infinite-time extinction probability $d_\infty(\epsilon)$ can be found by rewriting \cref{eq:dt_recursive} in the limit $t \to \infty$:
\begin{equation}
  d_\infty(\epsilon) = \min \left( 1, 
     \bigg(\frac 1 {\gamma(\epsilon)}-1 \bigg)^2\right) \ . 
\end{equation}

\begin{figure}
  \centering
  \includegraphics[width=0.4\textwidth]{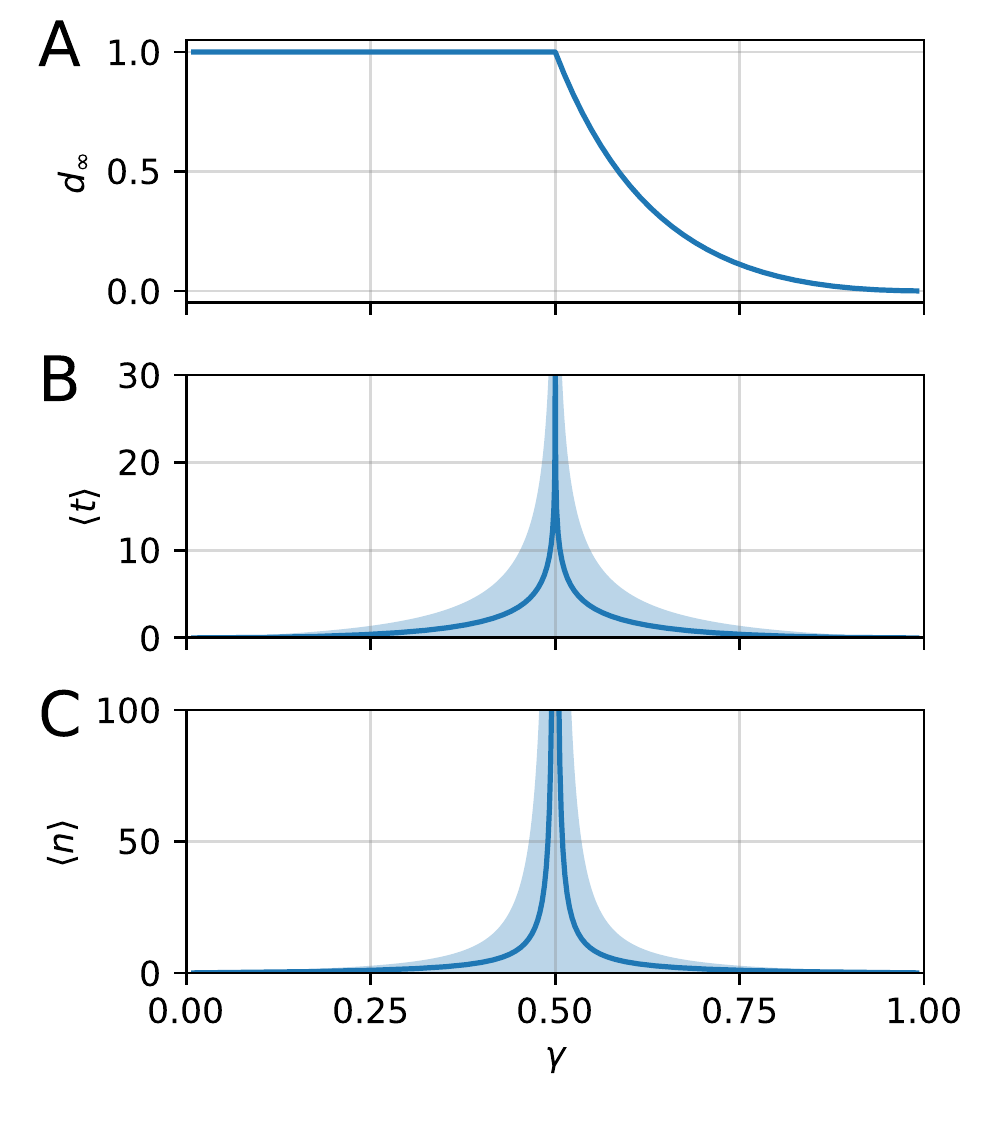}
  \caption{Value of the extinction probability $d_{\infty}$ (\textbf{A}), average extinction time $\avg{t}$ (\textbf{B}) and average progeny size $\avg{n}$ (\textbf{C}) as a function of the survival probability $\gamma(\epsilon)$ (cf. \cref{eq:gamma_nomut}) in the approximation of no affinity-affecting mutation. In B and C shaded area covers one standard deviation. Notice how extinction times and genealogy sizes diverge at $\gamma = \frac 12$.}
  \label{fig_app:crit_nomut}
\end{figure}

As expected, lineages will always go extinct if the average number of surviving offspring at division is not greater than one: $d_\infty(\epsilon) =1$ if $\gamma(\epsilon)\le \frac 12$. In \cref{fig_app:crit_nomut}A  we report the behavior of $d_{\infty}$ as a function of $\gamma$. This function presents a singularity at the critical value $\gamma = 1/2$, for which the Galton-Watson process is critical.

Finding an explicit expression for the distribution of extinction times is harder, but results can be obtained for the critical value $\gamma = \frac 12$. We find that the extinction time probability  behaves asymptotically as a power law, with infinite mean and variance: $r_t \sim 4/ t^{2}$ for large $t$. This result, which is a known feature of critical Galton-Watson processes, can be verified by inserting the Ansatz $d_t \sim 1 - \alpha t^{-1} + o(t^2)$ in \cref{eq:dt_recursive}, together with the simplified form of the mutation kernel (\ref{eq:mut_ker_nomut}) and the assumption that $\gamma = 1/2$. The only admissible solution is $\alpha = 4$ which, combined with the definition of $r_t$ \cref{eq:rt}, proves the statement. In \cref{fig_app:crit_nomut}B we display the mean and variance of the extinction time distribution as a function of $\gamma$. Comparison with \cref{fig:tree_stats}B shows that the divergence is removed when evolution includes affinity-affecting mutations. Mutations drive lineages away from the critical line, either to high affinities and  survival, or to lower affinities and extinction.

\subsection{Case of full population}
\label{subs:pop_ext_prob}

Building on the results derived above, we now turn to the problem of quantifying the average probability of extinction for the whole population.

\begin{figure*}[t]
    \centering
    \includegraphics[width=0.7\textwidth]{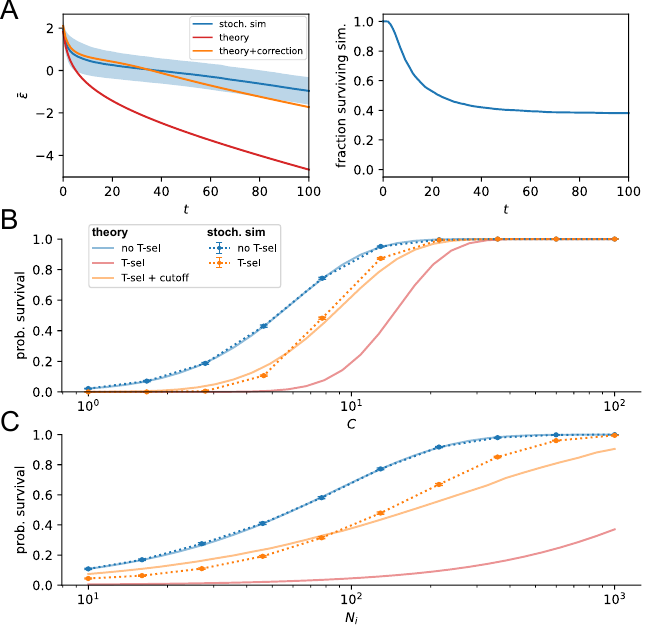}
    \caption{Probability of population survival in a bottleneck condition as a function of initial population size $N_i$ and Ag concentration $C$. \textbf{A}: Left: comparison between the evolution of $\bar\epsilon$ in stochastic simulations (blue, mean and standard deviation over 5000 simulations) and theoretical prediction without (red) and with (orange) finite-size correction. This correction consists in cutting the tail of the initial energy distribution in proximity of the expected value for the highest-affinity individual. The correction improves the prediction for the evolution of $\bar\epsilon$ at short times. Right: fraction of surviving simulations as a function of evolution round. With the finite-size correction the value of $\bar\epsilon$ is well-approximated during the time it takes for most of the simulations to go extinct. In this example we set $C=7$, $N_i = 100$. \textbf{B}: bottleneck survival probability as a function of antigen concentration $C$. Comparison between stochastic simulations (dotted line, error bars indicate the standard error of the mean) and the predictions our theory (full lines). Stochastic simulations are reported both without (blue) and with (orange) competitive selection for T-cell help (T-sel). For the theory instead we consider the case without T-sel (blue), with T-sel (red) and with T-sel plus finite-size correction (orange). In the absence of T-sel all cells evolve independently, and the theory and simulations match exactly. The inclusion of T-sel slightly decreases the survival probability in stochastic simulations. Accounting for this contribution by using the infinite-size estimate for the evolution of $\bar\epsilon$ overestimates the selection pressure. Adding the finite-size correction results in a much better estimate. In this example we set $N_i = 100$. \textbf{C}: same as B, but the survival probability is evaluated as a function of the initial size of the population $N_i$. Here we set $C=7$.}
    \label{fig:pop_psurv}
\end{figure*}

As a first approximation we do not consider competitive selection, since most of the selection pressure in a bottleneck is given by Ag-binding selection. Given a total of $N_i$ cells in the initial population, having energies $\{\epsilon_k\}_{k=1\ldots N_i}$ the probability that all cells will be extinct by evolution round $t$ is simply given by the product of extinction probabilities for all cells $\prod_k d_t (\epsilon_k)$. Moreover, since the initial energies are independently extracted from a Gaussian distribution $\varphi(\epsilon)$ with mean $\mu_i$ and standard deviation $\sigma_i$, the average extinction probability by round $t$ over all possible extractions of the initial population is given by:
\begin{equation}\label{eq:psurv_pop}
	P_\text{ext}(t) = \left(\int d\epsilon \; \varphi(\epsilon) \; d_t(\epsilon)\right)^{N_i}
\end{equation}
With the help of this formula we evaluate the average survival probability as a function of Ag concentration $C$ and initial population size $N_i$, and compare the prediction with stochastic simulations in which we turn off T-cell selection. The results, reported in \cref{fig:pop_psurv}B and C (blue), match exactly.

In the presence of competitive selection  the empirical survival probability evaluated from simulations slightly decreases, compare blue and orange dotted lines in \cref{fig:pop_psurv}B and C. The theory can be extended to account for T-selection in an effective manner. In practice, one needs first to extend the theory to include a time-dependence of the survival probability. At this point competitive selection can be included by introducing an effective coupling between cells in a `mean field' fashion, by estimating the average evolution of the term $\bar\epsilon = - \log \avg{e^{-\epsilon}}_\mathrm{pop}$ contained in the expression for the T-selection survival probability \cref{eq:psurv_t}.

Assume that the survival probability $P_S(\epsilon,t)$ is now time-dependent. The probability of extinction does not depend solely on the number of evolution rounds anymore, but also on the initial time at which the progenitor is considered. We define $d_{t,s}(\epsilon)$ as the probability that a cell, which at the end of round $t$ has binding energy $\epsilon$, will have all of its offspring extinct by the end of round $s > t$. For any value $t \geq 0$ we can write as before the probability of extinction in one round:
\begin{equation}
	\label{eq:d_first_BT}
	d_{t, t+1}(\epsilon) = \left(1 - \int d\Delta \, K(\Delta) \, P_S(\epsilon + \Delta, t) \, (1 - p_\mathrm{diff}) \right)^2
\end{equation}
And for any pair of rounds $s > t \geq 0$, with $s - t > 1$, the following recursive relation, analogous to \cref{eq:dt_recursive}, holds:
\begin{multline}
	\label{eq:d_iterative_BT}
	d_{t,s}(\epsilon) = \bigg[ 1 - \int d\Delta \, K(\Delta) \, P_S(\epsilon + \Delta, t) \, (1 - p_\mathrm{diff}) \\
	\times \, \left( 1 - d_{t+1,s}(\epsilon + \Delta)\right)\bigg]^2
\end{multline}
Finally, similarly to \cref{eq:psurv_pop}, the probability that the full population goes extinct by evolution round $t$ is given by:
\begin{equation} \label{eq:psurv_pop_BT}
	P_\text{ext}(t) = \left( \int d\epsilon \, \varphi(\epsilon) \, d_{0,t}(\epsilon)\right)^{N_i}\ ,
\end{equation}
where  $\varphi(\epsilon)$ is a Gaussian distribution with mean $\mu_i$ and standard deviation $\sigma_i$.

At this point we can make explicit the time dependence of the survival probability including selection for T-cell help: $P_S(\epsilon,t) = P_\mathrm{Ag} (\epsilon) \, P_\mathrm{T}(\epsilon, \bar\epsilon_t)$ (cf. \cref{eq:psurv_t}). Given the stochastic nature of our model, the variable $\bar\epsilon_t$ which quantifies selection pressure is in reality a stochastic variable. We estimate its average evolution using the large-population-size limit described in \cref{app:eb}, under which the model becomes deterministic. This allows us to numerically evaluate the extinction probability \cref{eq:psurv_pop_BT}. The outcome, however, underestimates the real survival probability (compare red curve and orange dotted line in \cref{fig:pop_psurv}B and C). This mismatch originates mainly from the fact that in the big-size approximation $\bar\epsilon$ evolves faster than in stochastic simulations (cf. blue and orange line in \cref{fig:pop_psurv}A-left). In turn, this occurs because the value of $\bar\epsilon$ is strongly dependent on the high-affinity tail of the population, whose evolution is influenced by finite-size effects.

This discrepancy can, however, be reduced with a simple finite-size correction. This correction is based on the consideration that the large-size limit of the model approximates the population binding energy histogram with a continuous distribution, encoded in the density function $\rho_t(\epsilon)$ (cf. \cref{app:eb}). At the beginning of evolution this function takes the shape of a normal distribution, corresponding to the initial binding energy distribution of naive responders, with tails extending indefinitely in both directions. As the population is finite in reality, consisting of $N_i$ individuals, we do not expect these tails to be populated. The correction procedure consists in removing these tails, by setting the initial distribution equal to zero outside a range delimited by two values $[\epsilon^-, \epsilon^+]$.

These two values are chosen equal to the expected energy of, respectively, the highest and lowest affinity individual in the population. The probability distribution for their binding energies can be expressed as a function of the naive binding energy distribution $\varphi(\epsilon)$ (as before a Gaussian with mean $\mu_i$ and variance $\sigma_i^2$) from which the energy of all cells is extracted. If we call $F(\epsilon) = \int_{-\infty}^\epsilon d\epsilon' \, \varphi(\epsilon')$ the cumulative distribution function, then these distributions can be expressed as:
\begin{align}
	\varphi^+(\epsilon) & = \frac{d}{d\epsilon}[F(\epsilon)]^{N_i}\\
	\varphi^-(\epsilon) & = -\frac{d}{d\epsilon} [1 - F(\epsilon)]^{N_i}
\end{align}
The values $\epsilon^\pm$ simply correspond to the means of these distributions.

Removing the tails to the initial distribution causes an initial slow-down in the evolution of $\bar\epsilon$ (cf. green line in \cref{fig:pop_psurv}A-left). This slow-down is eventually lost, but the agreement remains for a time sufficient for most of the stochastic simulations to go extinct (cf. \cref{fig:pop_psurv}A-right) which is the relevant timescale to capture bottleneck survival.

Taking the value of $\bar\epsilon$ obtained by combining the big-size approximation (cf. \cref{app:eb}) with the cutoff correction described above, and using it to evaluate the population survival probability, we obtain a much better agreement of the theory with simulations (compare orange curve and orange dotted line in \cref{fig:pop_psurv}B and C). The remaining discrepancy are due to the fact that the average evolution of $\bar\epsilon$ is still not exactly captured, and the `mean-field' nature of our approximation, which neglects the feedback of the energies in the population onto $\bar\epsilon$.

\section{Lineage size at extinction}
\label{subs:ext_size}

In this Section we focus on the distribution of sizes of the progeny at extinction. This size strongly depends on the model parameters, such as the  energy of the progenitor. An example is displayed in \cref{fig:explanation}A, in which  the lineage consists of a total of 82 cells. Like extinction time, this quantity is well-defined only for lineages that go extinct. Populations that are able to pass the bottleneck undergo exponential growth, with a rate that can be calculated from the large-size theory of Appendix~B, see \cite{molari2020elife}.

\subsubsection{Recursion equations for the distribution of sizes}

Similarly to what was done in the previous Section for the extinction time and probability, we now derive a recursive formula to quantify the total offspring size. We need to keep track of the sum of two random variables representing the numbers of descendants of each daughter cell. The recursion therefore includes a convolution, which is numerically harder to compute but can be handled using probability generating functions. The recursive relations can be expressed in term of these functions, and can be used to evaluate the moments of the probability distribution without having to numerically perform the convolution.

We name $q_n(\epsilon)$ the probability that a progenitor with energy $\epsilon$ generates a total offspring of exactly $n$ cells before extinction, not counting the progenitor itself (see \cref{fig:explanation}B). This probability can be better expressed if we separate the contribution of the two daughter cells. Considering genealogies encoded as binary trees, we call $b_m(\epsilon)$ the probability that along the branch corresponding to one of the daughter cells of a progenitor with energy $\epsilon$ we find a total of $m$ descendants (including the daughter cell itself) before extinction (see \cref{fig:explanation}). The expression for $m=0$ is simply given by the probability that the daughter cell is removed before the end of the round:
\begin{equation}
\label{eq:bzero}
\begin{split}
	b_0(\epsilon) &= 1 - \int d\Delta \, K(\Delta) \, P_S(\epsilon + \Delta) \, (1 - p_\mathrm{diff})\\
	 &= \sqrt{d_1(\epsilon)}
\end{split}
\end{equation}
The recursive relation in this case is composed of two equations. The first is a convolution that decomposes the probability of having $n$ descendants as a sum over all possible repartitions of the descendant number along the two branches:
\begin{equation}\label{eq:qn_convoltion}
	q_n(\epsilon) = \sum_{m=0}^n b_m (\epsilon) \; b_{n-m}(\epsilon)
\end{equation}
The second expresses the probability to find $m$ descendants along a branch as the probability that the daughter cell survives and has $m-1$ descendants:
\begin{equation}\label{eq:qn_recursive}
	b_m(\epsilon) = \int d\Delta \, K(\Delta) \, P_S(\epsilon + \Delta) \, (1 - p_\mathrm{diff}) \, q_{m-1}(\epsilon + \Delta)
\end{equation}
We introduce the generating functions $Q(z, \epsilon)$ and $B(z,\epsilon)$, defined as:
\begin{equation}\label{eq:QB_def}
    Q(z, \epsilon) = \sum_{n = 0}^\infty q_n(\epsilon) \, z^n, \qquad B(z, \epsilon) = \sum_{m = 0}^\infty b_m(\epsilon) \, z^m \ .
\end{equation}
In terms of these generating functions equations \cref{eq:qn_convoltion,eq:qn_recursive} become
\begin{gather}
  \label{eq:Q_conv}
  Q(z, \epsilon) = B(z, \epsilon)^2\\
  \label{eq:Q_rec}
  \begin{split}
  	\frac{1}{z} [B(z, \epsilon) - b_0(\epsilon)] = \int d\Delta \, K(\Delta) \, P_S(\epsilon + \Delta) \\
  \times \, (1 - p_\mathrm{diff}) \, Q(z, \epsilon + \Delta)
  \end{split}
\end{gather}
These relations can be used to evaluate the moments of these distributions with two additional considerations. The first is that $\sum_{n=0}^{\infty} q_n(\epsilon) = d_\infty (\epsilon)$. This sum does not converge to one since it only considers lineages that eventually go extinct. For the functions $Q$ and $B$ this translates into:
\begin{equation}
    Q(z=1, \epsilon) = d_{\infty}(\epsilon), \qquad B(z=1, \epsilon) = \sqrt{d_{\infty}(\epsilon)}\ .
\end{equation}
Secondly, the moments of the distributions can be evaluated from the generating functions as:
\begin{gather}
\label{eq:n_moments}
\begin{split}
	\avg{n^k}_{\epsilon} &= \frac{1}{d_{\infty}(\epsilon)} \, \sum_{n = 0}^{\infty} n^k \, q_n(\epsilon) \\
	&= \frac{1}{d_{\infty}(\epsilon)} \, \left(z \partial_z \right)^k \, Q(z,\epsilon) |_{z=1}
\end{split}\\
\begin{split}
  \avg{m^k}_{\epsilon} &= \frac{1}{\sqrt{d_{\infty}(\epsilon)}} \, \sum_{m = 0}^{\infty} m^k \, b_m(\epsilon) \\
  &= \frac{1}{\sqrt{d_{\infty}(\epsilon)}} \, \left(z \partial_z \right)^k \, B(z,\epsilon) |_{z=1}
\end{split}
\end{gather}
Applying the operator $z \partial_{z}$ one and two times on \cref{eq:Q_conv} restitutes the following relations between the first two moments:
\begin{equation} \label{eq:n_m_moments}
    \avg{n}_{\epsilon} = 2\avg{m}_{\epsilon} \, , \qquad
    \avg{n^2}_{\epsilon} = 2 \avg{m^2}_{\epsilon}+ 2\avg{m}^2_{\epsilon}
\end{equation}
This corresponds simply to the fact that the total number of descendants is the sum of the descendants along the two branches. Applying the same operator on \cref{eq:Q_rec} gives:
\begin{gather}
	\label{eq:m_firstmom}
	\begin{split}
	\sqrt{d_{\infty}(\epsilon)} \, \avg{m}_{\epsilon} = \int d\Delta \, K(\Delta) \, P_S(\epsilon + \Delta) \, (1 - p_\mathrm{diff}) \\
	\times \, d_{\infty}(\epsilon + \Delta) \, [2 \avg{m}_{\epsilon + \Delta} + 1]
	\end{split}\\
	\label{eq:m_secondmom}
	\begin{split}
		\sqrt{d_{\infty}(\epsilon)} \, \avg{(m-1)^2}_{\epsilon} = \int d\Delta \, K(\Delta) \, P_S(\epsilon + \Delta) \\
		\times \, (1 - p_\mathrm{diff}) \, d_{\infty}(\epsilon + \Delta) \, [2 \avg{m^2}_{\epsilon + \Delta}+ 2\avg{m}^2_{\epsilon + \Delta} + 1]
	\end{split}
\end{gather}
These equations can be solved numerically if we express them as fixed-point equations for the functions $\avg{m}_\epsilon$ and $\avg{m^2}_\epsilon$:
\begin{gather}
\begin{split}
	\avg{m}_{\epsilon} = \frac{1}{\sqrt{d_{\infty}(\epsilon)}} \int d\Delta \, K(\Delta) \, P_S(\epsilon + \Delta) \, (1 - p_\mathrm{diff}) \\ \times \, d_{\infty}(\epsilon + \Delta) \, [2 \avg{m}_{\epsilon + \Delta} + 1]
\end{split}\\
	\begin{split}
		\avg{m^2}_{\epsilon} = \frac{1}{\sqrt{d_{\infty}(\epsilon)}} \int d\Delta \, K(\Delta) \, P_S(\epsilon + \Delta) \, (1 - p_\mathrm{diff})  \\ 
		\times \, d_{\infty}(\epsilon + \Delta) \, [2 \avg{m^2}_{\epsilon + \Delta} + 2 \avg{m}^2_{\epsilon + \Delta} + 4 \avg{m}_{\epsilon + \Delta} + 1]
	\end{split}
\end{gather}
The moments for $n$ can then easily be evaluated using \cref{eq:n_m_moments}.

In \cref{fig:tree_stats}C we compare the theoretical prediction for the first two moments (orange line represents the mean and shaded area covers one standard deviation) with the corresponding quantities from stochastic simulations (blue, error bars cover one standard deviation). Once more we find a good match. The peak at intermediate affinities can be explained, as done above for the extinction time, considering the critical nature of this phenomenon at intermediate values of the binding energy. This is done in the next section.

\subsubsection{Exactly solvable case of no mutation}

Similarly to what done for extinction probability, in the absence of affinity-affecting mutations we can find an explicit expression for the mean and variance of the population size at extinction. By plugging the simplified expression for the mutation kernel \cref{eq:mut_ker_nomut} into \cref{eq:Q_conv,eq:Q_rec} one obtains the following second degree equation for the generating function $B$:
\begin{equation} \label{eq:B_eq_nomut}
    z \, \gamma(\epsilon) \, B(z,\epsilon)^2 - B(z,\epsilon) + 1-\gamma(\epsilon) = 0
\end{equation}
Where as before $\gamma(\epsilon)$ is the probability for a daughter cell not to be removed from the population during the evolution round, cf. \cref{eq:gamma_nomut}. This equation has two solutions. The correct one can be chosen by considering that $B$ must be a monotonically increasing function of $z$. This gives:
\begin{equation}\label{eq:B_nomut}
    B(z, \epsilon) = \frac{1 - \sqrt{1 - 4 \, z \, \gamma(\epsilon) \, (1-\gamma(\epsilon))}}{2 \, z \, \gamma(\epsilon)}
\end{equation}
The function $Q(z,\epsilon)$ can be evaluated from \cref{eq:Q_conv}, and the mean and variance for the population extinction sizes can be obtained using \cref{eq:n_moments}. This results in:
\begin{align}
    \avg{n}_{\epsilon} &= \begin{cases}
    \frac{2 \gamma(\epsilon)}{1 - 2 \gamma(\epsilon)} & \text{if } \gamma(\epsilon) < 1/2\\
    \frac{2 - 2 \gamma(\epsilon)}{2 \gamma(\epsilon) - 1}& \text{if } \gamma(\epsilon) > 1/2
    \end{cases}\\
    \avg{n^2}_\epsilon - \avg{n}^2_\epsilon &= \avg{n}_\epsilon (\avg{n}_\epsilon + 1) (\avg{n}_\epsilon + 2)/2 
\end{align}
This quantity are reported as a function of $\gamma$ in \cref{fig_app:crit_nomut}C. Similarly to what observed for the mean and variance of the extinction time, Both of these quantities diverge for the critical value $\gamma = \frac 12$, but this divergence is removed when mutations are considered.

It is interesting to consider the effect of this divergence on the coefficients $q_n(\epsilon)$, that represent the probability of a lineage that stems from a progenitor with binding energy $\epsilon$ to go extinct with a total progeny of $n$ cells (see \cref{fig:explanation}B). From the definition of the generating function $Q(z,\epsilon)$ (cf. \cref{eq:QB_def}) it follows that the coefficients $q_n$ can be obtained by Taylor expansion of this function around $z = 0$. In turn $Q$ can be easily obtained from \cref{eq:B_nomut} using the property $Q = B^2$ (see \cref{eq:Q_conv}). The expansion results in the following expression for the coefficients when mutations are absent:
\begin{eqnarray}
\label{eq:qn_nomut}
    q_n &=& \frac{(2n+2)!}{(n+1)!(n+2)!} \gamma^n (1-\gamma)^{n+2} \nonumber \\
    &\stackrel{n \gg 1}{\sim}& \frac{(1-\gamma)^2}{\sqrt{\pi n^3}}(4 \gamma (1 - \gamma))^n
\end{eqnarray}
In general these probabilities decay exponentially fast as a function of the size $n$. At the critical value $\gamma=1/2$ however the term $4 \gamma (1 - \gamma)$ becomes equal to $1$ and the coefficients algebraically decay as $n^{-3/2}$. The asymptotic decay of the coefficients $q_n$ for $n \gg 1$ can also be obtained from the behavior of the generating function $Q$ around its singularity at $z_c = 1 / (4 \gamma (1-\gamma))$. In particular  $Q \propto (1-z/z_c)^{1/2}$ close to the singularity, and therefore $q_n \propto n^{-3/2} \, z_c^{-m}$ \cite{flajolet2009analytic}, which gives the expected asymptotic behavior described above.

\subsubsection{Case of small-effect mutations}

Corrections to the asymptotic behavior above arise when small-effect mutations are considered. In particular in \cref{eq:mut_ker_nomut} we substitute the Dirac delta distribution with a peaked Gaussian, and consider a mutation kernel having the form:
\begin{equation}\label{eq:Kmut_weak}
    K(\Delta) = (1 - p_\mathrm{let}) \; \frac{1}{\sqrt{2\pi}\sigma}\exp \left(-\frac{\Delta^2}{2 \sigma^2}\right) \quad \text{with } \sigma \ll 1
\end{equation}
When the standard deviation is small enough one can approximate the integrals in \cref{eq:bzero,eq:Q_rec} by Taylor-expanding the functions that multiply the mutation kernel around $\Delta = 0$. This results in the following approximation for \cref{eq:Q_rec}:
\begin{equation}
    B - 1 + \gamma + \frac{\sigma^2}{2} \gamma'' = z\gamma B^2 + z\frac{\sigma^2}{2} \left[\gamma B^2\right]'' \ ,
\end{equation}
according to the definition of $\gamma$ (cf. \cref{eq:gamma_nomut}), and with inverted commas indicating derivatives with respect to $\epsilon$. This equation is analogous to \cref{eq:B_eq_nomut} with the addition of perturbation terms of the order of $\sigma^2$. We therefore suppose $B(z, \epsilon) \sim B_0(z,\epsilon) + \sigma^2 \Delta B (z,\epsilon)$ where $B_0$ is given by \cref{eq:B_nomut} and $\Delta B$ represents a perturbation to this $\sigma^2=0$ solution. By plugging this Ansatz in the previous equation we find the following expression for the perturbation: 
\begin{equation}\label{eq:perturbation}
    \Delta B = \frac{(B_0 - 1)}{2 \gamma r^4} \left[ 2 z \gamma'^2 (1 - 2 \gamma) + r\frac{\gamma'^2}{\gamma} + r^2 \left(\gamma'' - \frac{\gamma'^2}{\gamma}\right)\right],
\end{equation}
where $r$ is defined through
\begin{equation}
    r=\sqrt{1 - 4 z \gamma (1-\gamma)} = \sqrt{\frac{z_c-z}{z_c}} \ .
\end{equation}
The critical value of $z$ is given as before by $z_c = 1 / (4 \gamma (1-\gamma))$. Moreover, when the survival probability $P_S$ is given by \cref{eq:psurv_ag},  the derivatives of  $\gamma$ (cf. \cref{eq:gamma_nomut}) can be expressed as
\begin{align}
    \gm' &= -\gm (1-\wt\gm)\ . \\
    \gm'' &= \gm (1-\wt\gm) (1-2\wt\gm)\ ,
\end{align}
where $\wt\gm = P_\mathrm{Ag}(\epsilon)$.
From the expression for $\Delta B$ we can derive the perturbation to $Q$ by using \cref{eq:Q_conv}. Keeping only the higher order in $\sigma^2$, we obtain
\begin{equation}
    Q \sim Q_0 + \sigma^2 \Delta Q \, , \text{ with } \Delta Q = 2 \, B_0 \, \Delta B
\end{equation}
As stated in the previous section, the behavior of the probabilities $q_n$ is strictly related to the behavior of $Q$ around its singularity $z_c$. In particular the magnitude of the perturbation to the coefficients $q_n$ introduced by mutations can be derived from the study of $\Delta Q$. If we operate the substitution $z = z_c (1 - r^2)$ and expand $\Delta Q$ in powers of $r$ we obtain:
\begin{equation}\label{eq:deltaQ_powers}
    \Delta Q = c_4 r^{-4} + c_3 r^{-3} + c_2 r^{-2} + c_1 r^{-1} +O(1)
\end{equation}
with the following expressions for the coefficients:
\begin{equation}
\begin{split}
c_4 &= (1 - 2 \gm)^2 (1 - \wt \gm)^2\\
c_3 &= -(1 - 2 \gm)^2 (1 - \wt \gm)^2\\
c_2 &= - 2 (1 - \gm) (1 - \wt \gm)(1 - 2 \gm \wt \gm)\\
c_1 &= 2 (1 - \gm) (1 - \wt \gm)(3 - 2 \gm - 2 \gm\wt \gm)
\end{split}
\end{equation}
Based on this expansion  we can derive an expression for the perturbation to the coefficients $\Delta q_n = q_n - q_n^0$ caused by weak mutations  \cite{flajolet2009analytic}, where $q_n^0$ represents the value in the absence of mutations (cf. \cref{eq:qn_nomut}). We obtain that, for large $n$, 
\begin{multline}\label{delqn}
    \Delta q_n = \sigma^2 [c_4(n+1) +  c_3 (n/\pi)^{1/2}(2 + 3/4n) + c_2 \\+ c1 (\pi n)^{-1/2} + O(n^{-3/2})](4 \gamma (1-\gamma))^n
\end{multline}

\begin{figure*}
    \centering
    \includegraphics[width=0.8\textwidth]{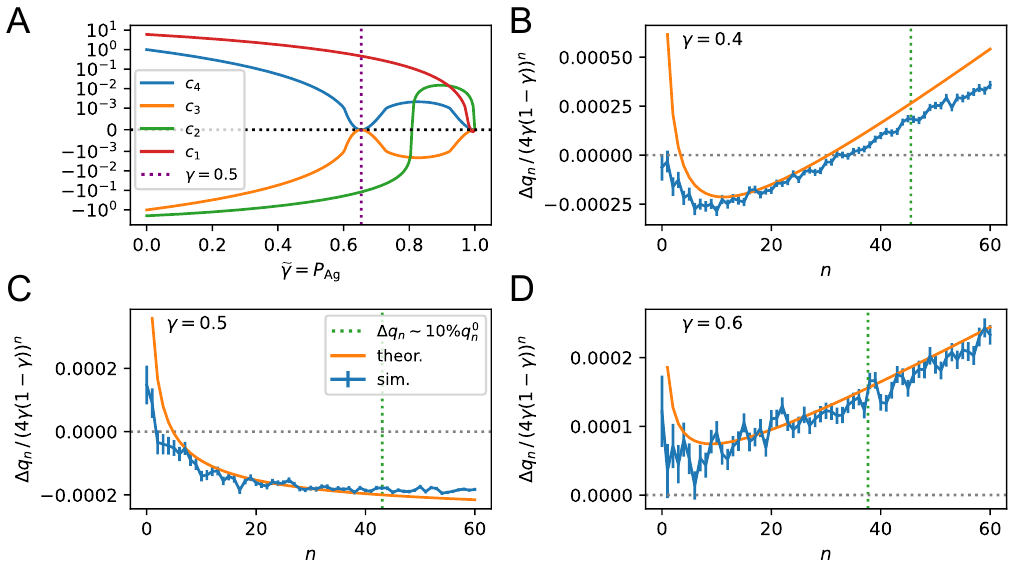}
    \caption{\textbf{A}: values of the power expansion coefficients in \cref{eq:deltaQ_powers} as functions of the progenitor survival probability $\wt\gm = P_\mathrm{Ag}$. Both $c_4$ and $c_3$ are null at the critical value $\gm=1/2$. \textbf{B,C,D}: perturbations to the probabilities $q_n$ due to weak mutations (cf. \cref{eq:Kmut_weak}, with $\sigma = 0.05$) for three different progenitor affinities, corresponding to $\gm = 0.4, \, 0.5,\, 0.6$ (B,C and D respectively). The perturbation $\Delta q_n = q_n - q_n^0$ is evaluated as the difference between the probability of lineage extinction at progeny size $n$ without ($q_n^0$) and with ($q_n$) mutations. These were evaluated by performing $10^8$ numeric simulations for each of the two conditions. In the plots we report the value $\Delta q_n \, (4 \gm (1-\gm))^{-n}$ from simulations (blue) and compare it with the theoretical prediction (orange). The value of $n$ at which $|\Delta q_n| = 10\% \, |q_n^0|$ sets an upper limit for the validity of the theory (green).}
    \label{fig:qn_correction}
\end{figure*}

In \cref{fig:qn_correction}A we plot the values of the coefficients $c$ as a function of $\wt\gm$. It is interesting to notice that both $c_4$ and $c_3$, controlling the two leading orders, are null for the critical value $\gamma=1/2$. This leaves the next leading order to $c_2$ which, for this value of $\gm$, is negative. As a result, we expect that the perturbation tends to lower the values of $q_n$ at large $n$ for $\gamma\sim 1/2$, and to raise it for $\gamma \lessgtr 1/2$. This means that large-size extinction events are made less probable by the presence of mutations around the critical value, which is consistent with the removal of the divergence observed in \cref{fig_app:crit_nomut}C when mutations are present.

In \cref{fig:qn_correction}B,C,D we compare the above prediction for $\Delta q_n$ with the value provided by numerical simulations for different values of $\gm$, as indicated in each plot. We expect the theoretical prediction to be accurate for large values of $n$, and as long as the perturbation $\Delta q_n$ remains small with respect to the unperturbed value $q^0_n$. However, for increasing values of $n$ the perturbation grows faster than the unperturbed value, eventually invalidating this assumption. As a proxy for an accuracy upper limit we mark with a green dotted line the value of $n$ at which the perturbation $\Delta q$ has a magnitude equal to $10\%$ of the unperturbed value $q_n^0$. Based on eqn.~(\ref{delqn}) this value scales as $\sigma^{-4/5}$ for small $\sigma$'s.

\section{Discussion}
\label{sec:discussion}

In this work we focused on the effects of a bottleneck on a B-cell population in the course of the affinity maturation process. 
Through a recursive relation that links the probability of bottleneck survival of a cell to the one of its daughter cells we were able to retrieve the dependence of a lineage extinction probability on its progenitor affinity. For lineages that go extinct we also evaluated the mean and variance of extinction time and progeny size, revealing a peak in extinction time corresponding to average affinity progenitors. Lineages stemming from these progenitors spawn in equilibrium between extinction and survival, and persist in this state until mutations drive the lineage either to survival or to extinction. Building on these results we then evaluated the survival probability for the full population as a function of Ag concentration and population size. We also included the effect of competition in an effective manner, using the deterministic model limit combined with a finite-size correction.

The bottleneck phenomenology was included in different maturation models \cite{wang2015manipulating, zhang2010optimality}, which considered as optimal the maturation regime in which the B-cell population was subject to a strong enough selection force to grant good affinity enhancement, while at the same time not strong enough to cause population extinction. While the properties of the above models were numerically evaluated using stochastic simulations, we here present exact or approximate derivations for various quantities of interest through the use of recursive relations and probability generating functions. 
These techniques are often encountered in the context of branching processes, and similar approaches have been used in the study of AM \cite{balelli2018random,balelli2019multi}. In our case we coupled the theoretical analysis with a more realistic model that included both the effect of mutations and selection for Ag binding and competition. We believe our approach provides a better understanding of what controls the lineage survival probability, while at the same time requiring less computational resources when compared to averaging over many stochastic simulations. This allows one to easily explore the effect of changing different model parameters on the survival probability and the lineage size.

In this last section of the paper we discuss the biological relevance of our results, together with some perspectives. We focus on four aspects: the role of population bottleneck as a limiting case in biologically observed maturation, the affinity of the initial B-cell population, how our results could be applied to explain the role of precursor frequency and affinity on the successful colonization of GCs, and finally how the theory might also offer insight in the case of multiple antigens.\\

\subsection{Population bottleneck in affinity maturation}
\label{subs:bottleneck_bio}

\begin{figure*}[t]
    \centering
    \includegraphics[width=0.8\textwidth]{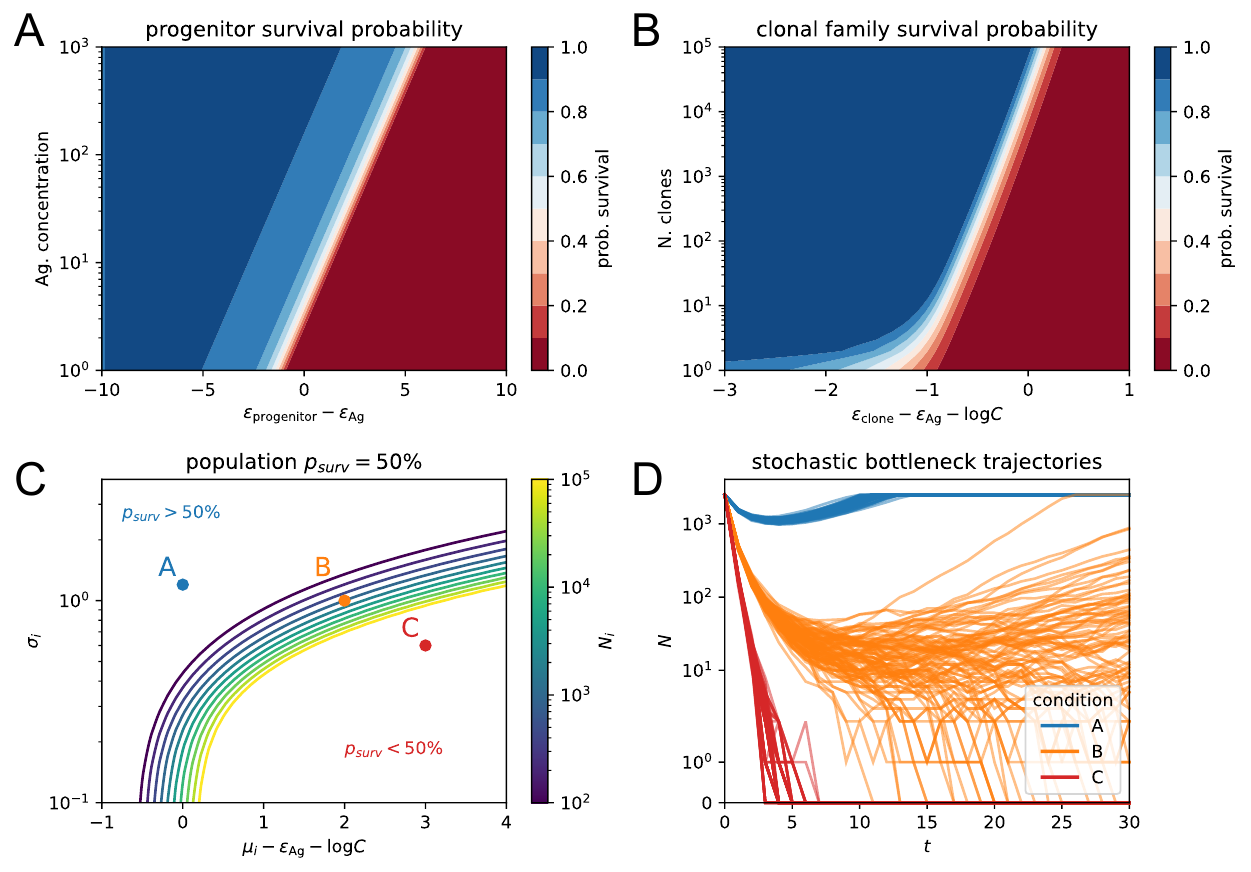}
    \caption{
    \textbf{A}: bottleneck survival probability of a lineage stemming from a single progenitor cell with energy $\epsilon_\mathrm{progenitor}$, as a function of antigen concentration. 
    \textbf{B}: probability of bottleneck survival for an initial population composed of $N$ identical cells with binding energy $\epsilon_\mathrm{clone}$.
    \textbf{C}: population survival probability as a function of its initial size $N_i$ and the mean $\mu_i$ and standard deviation $\sigma_i$ of the initial binding energy distribution. Colored lines indicate the region of the phase diagram in which the probability of survival is equal to 50\%. Different colors distinguish different initial population sizes according to the colorbar on the right. Each line divides the plane in two regions, corresponding to high ($p_\mathrm{surv} > 50\%$) and low ($p_\mathrm{surv} < 50\%$) survival probability. We consider three conditions that belong to either one of these two regions (condition A and C), or to the separation line (condition B) to be visualized in the next panel.
    \textbf{D}: evolution of total population size $N$ for stochastic model simulations realized under three different conditions as indicated in panel C. For each condition we display 100 different stochastic simulation trajectories, with time $t$ measured in evolution rounds. The simulation were performed at an initial population size $N_i = 2500$ and antigen concentration $C = 10$, while the mean and standard deviation $(\mu_i,\sigma_i)$ for the initial binding energy distribution were varied according to the condition (A: (2.3,1.2), B: (4.3, 1.), C: (5.3, 0.6)). Results displayed in panels A,B and C do not take into account competitive selection, while in panel D competitive selection is included in simulations.
    }
    \label{fig:psurv_parameters}
\end{figure*}

The population bottleneck phenomenon, intended as a transitory state of low population with a high extinction risk, is a common feature of many maturation models \cite{wang2015manipulating, zhang2010optimality}. However to our knowledge this phenomenon lacks experimental confirmation. This might be due to two main factors: experimental limitations on one hand, and biologically relevant conditions on the other.\\
By nature the experimental observations of germinal centers tends to be destructive: in order to study the cellular composition of a germinal center the animal must often be sacrificed. This allows for the contemporary observation of multiple germinal centers at the same point in time, but not for the repeated observation of the same germinal center at different time points. As a result it is difficult to estimate for example whether chronic germinal center reactions feature long-lived germinal centers, or many short-lived germinal centers that appear and wane \cite{victora2018primary}. To our knowledge the only system capable of circumventing these difficulties is the one introduced by Firl and colleagues \cite{firl2018capturing}, who developed an intravital imaging system that allows for observation of germinal centers for an extended period of time. Interestingly, they report that in around 40\% of their observations the germinal center failed to form, with the population of founder cells initially expanding until around 5 days after antigen administration, and then starting to wane. The authors observe that this might either be due to a competition with a non-imaged clone, or to a failure to establish a productive germinal center, which would be compatible with a "failed bottleneck" scenario. In short, while lacking explicit experimental confirmation, the bottleneck phenomenology is also not ruled out by experimental observations.\\
A second important consideration is that the bottleneck scenario might only represent a particular possible regime in maturation, which might be relatively rare under biologically relevant conditions. To illustrate this we consider the probability for a progenitor with energy $\epsilon_\mathrm{progenitor}$ to survive the bottleneck state, as a function of the amount of available antigen concentration and of the energy difference $\epsilon_\mathrm{progenitor} - \epsilon_\mathrm{Ag}$ (cf. \cref{fig:psurv_parameters}A). This can be easily evaluated using the $t \to \infty$ limit of recursion \cref{eq:dt_recursive}. Here we only consider Ag-binding selection as the major source of selective pressure in the bottleneck and neglect the effect of competition for T-cell help. One can observe a steep transition between a region of almost certain extinction (red) to a region of possible survival (white and blue). The width of the region is essentially controlled by the probability of developing a beneficial mutation, and the magnitude of such mutations. As evident from the form of the selection survival probability \cref{eq:psurv_ag} operating a multiplicative change of antigen concentration by a factor $\Delta C$ has the same effect as considering a change of binding energy of magnitude $-\log \Delta C$. One could therefore simply include the effect of concentration by rescaling the energy values. As a next step we consider the probability of survival for a full population, as a function of the initial population size $N_i$ and the mean $\mu_i$ and standard deviation $\sigma_i$ of the initial binding energy distribution of cells. In \cref{fig:psurv_parameters}C we display the values at which the population survival probability $p_\mathrm{surv}$ is equal to 50\%. These correspond to lines in the $\mu_i$ / $\sigma_i$ plane, with different colors corresponding to different initial population sizes $N_i$. The lines split the diagram in two regions, of high and low survival probability ($p_\mathrm{surv} \lessgtr 50\%$). To visualize the behavior of the system we consider three different cases, of high (case A), intermediate (case B) or low (case C) survival probability. In \cref{fig:psurv_parameters}D for each of these conditions we display the evolution of the total population size for 100 different stochastic simulations. In the low-survival-probability region, corresponding to case C, in all of the simulations the population quickly goes extinct. Conversely, in the high-survival-probability region (case A), the population only undergoes a partial reduction in size, with all of the simulations successfully surpassing the bottleneck. In the intermediate region (case C) we observe a consistent reduction in size for all of the simulations, with both successes and failures in overcoming the bottleneck. This intermediate region corresponds indeed to the bottleneck phenomenology that we are interested in studying in this paper, but biologically might only represent a subset of all the possible maturation regimes. In particular one might expect maturation to normally occur in the high-survival-probability regime in most of the biologically relevant cases, especially for pathogens that are similar to ones already encountered in the past for which high-affinity precursor might already be available in the initial population. The low-survival-probability regime conversely might never occur. In this case in fact no precursor with sufficient affinity is expected to be present in the initial population, and GCs might fail to form altogether. Finally the intermediate regime, in which the bottleneck phenomenology is observed, might be a limit case and only occur for ``hard'' antigens, when only few precursor of acceptable affinity exist in the initial population. These limit cases might be of particular interest when developing novel vaccination strategies, for example against mutable pathogens such as HIV, as will be discussed in \cref{subs:aff_prec}.\\

\subsection{Affinity of the initial population}

As shown in the previous section, the bottleneck phenomenology occurs only for values of the parameters that place the system in the boundary between the high and low survival probability region. This is fundamentally controlled by the affinity of the initial population: if enough high-affinity precursors are present in the initial population then the population survival is almost certain, while if all cells in the population have low affinity then survival is rare event. To study this phenomenology in our simulations we chose the size and affinity distribution of the initial population so as to be on this boundary region, while at the same time being compatible with experiments. This was done as follows.\\
The initial population size was set equal $N_i = 2500$. This is in agreement with \cite{eisen2014affinity} which reports around 3000 cells per germinal center. For simplicity we consider the binding energy of each cell to be independently extracted from a normal distribution with mean $\mu_i$ and standard deviation $\sigma_i$. This simplification overestimates the heterogeneity of the initial population binding energy, which in reality is usually composed of $50 \sim 200$ different clones \cite{tas2016visualizing}. The value of $\sigma_i = 1.5$ was chosen in agreement with \cite{molari2020elife}, in which a fit of experimental single-cell affinity measurements of naive responder cells results in an estimated $\sigma_i \sim 1.66$. Given these conditions the value of $\mu_i$ (together with antigen concentration $C$) controls the probability of surviving the bottleneck (cf. \cref{fig:psurv_parameters}C). For most of the simulation we choose a value $\mu_i = 4$, so that only around $2\%$ of cells from the initial population have affinity higher than the Ag-binding selection threshold for values of $C \sim 5$. This makes so that for standard parameter values the model is close to the phase threshold line (i.e. condition B in \cref{fig:psurv_parameters}C and D), which allows us to visualize and study the bottleneck phenomenology.\\
Under these assumptions the initial population will be composed in large part by low-affinity cells. In general, while antigen affinity does indeed drive the recruitment of cells in germinal centers, the presence of cells with undetectable affinity has been observed in different experiments \cite{abbott2020factors}. For example in the experiments reported in \cite{eyer2020quantitative} only a minor fraction ($\sim 30\%$) of all the antibody-secreting cells elicited by immunization was showing detectable affinity for the administered antigen ($K_d <$ 500 nM). Moreover, as experimentally shown in \cite{schwickert2011dynamic}, in the absence of competition even cells with very low affinity for the antigen ($K_d \sim 8 \; \mu$M) can colonize GCs. The role of low-affinity cells in maturation has not yet been experimentally elucidated. Their presence has either been attributed to some form of non-specific activation (e.g. bystander activation \cite{inoue2018generation,horns2020memory}), or to the specific binding of some unknown ``dark antigen'' \cite{abbott2020factors}, or again to specific binding of the administered antigen but with a very low (undetectable) affinity. In either case their number usually decreases during maturation, in favor of higher affinity cells.\\

\subsection{Effect of affinity and precursor frequency on germinal center colonization}
\label{subs:aff_prec}

The theory we introduced in this paper offers a framework to quantify the probability for a cell lineage to be able to survive the population bottleneck, successfully colonize a GC and undergo maturation. This has strong connections with the field of vaccine design. In this field the development of new and effective vaccination techniques against pathogens such as HIV requires the stimulation of germline clones that are oftentimes rare or have low affinity for the target. To design an effective vaccine it is indeed important to understand the effect of germline affinity and precursor frequency on the successful recruitment of clones in germinal centers. Recent experiments have shown that at physiological levels of affinity and frequency, successful recruitment depends in a non-trivial way on both of these variables \cite{abbott2020factors, abbott2018precursor,havenar2018designing}.\\
In this context, our theory could offer a quantitative prediction for the probability of a clonal family to survive selection, as a function of the initial abundance of cells and of their affinity. To illustrate this we consider a group of $N$ identical cells with initial energy $\epsilon_\mathrm{clone}$. Using our theory we can evaluate the probability that the offspring of at least one cell will survive the population bottleneck. This probability is displayed in \cref{fig:psurv_parameters}B. In agreement with what observed in experiments \cite{abbott2020factors, abbott2018precursor}, successful survival depends on both $N$ and $\epsilon_\mathrm{clone}$, with survival probability being increased by both higher affinity and higher precursor frequency. A quantitative understanding of this trade-off might guide decisions in the development of vaccine antigens and improve experiment design.\\

\subsection{Case of multiple antigens}

\begin{figure*}[t]
    \centering
    \includegraphics[width=0.65\textwidth]{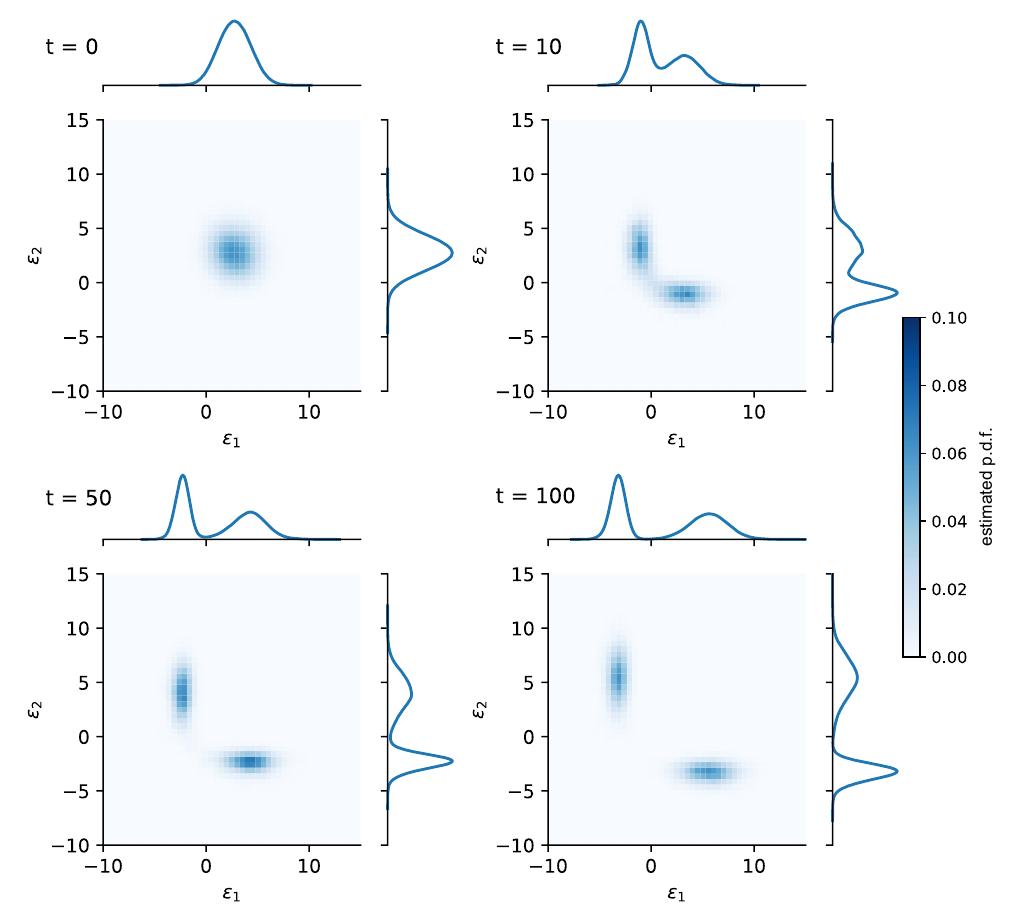}
    \caption{
    Joint probability distribution $P_t(\epsilon_1, \epsilon_2)$ of the binding energies ($\epsilon_1$, $\epsilon_2$) of the B-cell population in the two-antigens case described in the text. The four different plots correspond to four different times ($t = 0, 10, 50, 100$), measured in evolution rounds. The probability distribution was estimated by averaging over 500 different stochastic simulations. For each plot we also display the marginal probabilities $P_t(\epsilon_1)$ and $P_t(\epsilon_2)$. Simulations were performed at a value of antigen concentration $C_1=C_2=5$, and considering an initial average binding energy $\mu_i = 3$ for both antigens.
    }
    \label{fig:mags}
\end{figure*}

It would be important to extend the present work to the case of multiple antigens. Contrary to the case of simple antigens, where the response is usually focused on a single dominating epitope, in the case of complex or multiple antigens there might be competition between the binding of different epitopes. Understanding how maturation plays out in the presence of multiple antogens is currently an open issue, whose solution could help the development of vaccination strategies against mutable pathogens such as HIV \cite{chakraborty2017rational, victora2018primary, robert2018induction}.\\
As a natural extension to our theory one could consider that a single cell might possess different affinities for each antigen mutant. These affinities are potentially correlated, depending on the similarity between the different mutants, and so is the effect of mutations. Taking this into account one could then define mutation and selection probabilities in the presence of multiple Ag mutants, and use a similar approach to the one introduced in this paper to evaluate the lineage and population survival probability.\\
Even without such extension however we believe that the single-antigen framework might still be instructive in the study of multiple antigen maturation. To demonstrate this one can extend the model to the case of two antigens, in a similar fashion to what was done in \cite{wang2015manipulating}. In this extension each cell possesses two binding energies $\epsilon_1$ and $\epsilon_2$, encoding its affinity for the two antigens. The values of these energies are initially independently extracted from the same normal distribution of "naive" binding energies. The system then evolves similarly to the single-antigen case, with two differences. The first one is that affinity-affecting mutations change the binding energy of the cell for both antigens with two random contributions $\Delta \epsilon_1$ and $\Delta \epsilon_2$, independently extracted from the usual mutation kernel (cf. \cref{eq:mut_ker}). Notice that as a result of this, mutations that increase the affinity for only one of the two antigens are much more common than mutations that increase the affinity for both. The second difference is in the way selection is performed. Following the ``meet-all'' scenario introduced in \cite{wang2015manipulating} we extend the definition of the selection survival probabilities as:
\begin{gather}
    P_\text{Ag}(\epsilon_1, \epsilon_2) = \frac{C_1 e^{-\epsilon_1} + C_2 e^{-\epsilon_2}}{C_1 e^{-\epsilon_1} + C_2 e^{-\epsilon_2} + e^{-\epsilon_\text{Ag}}}\\
    \begin{split}
        P_\text{T}(\epsilon_1, \epsilon_2) = \frac{C_1 e^{-\epsilon_1} + C_2 e^{-\epsilon_2}}{C_1 e^{-\epsilon_1} + C_2 e^{-\epsilon_2} + e^{-\bar\epsilon}},\\
        \mathrm{with} \quad e^{-\epsilon} = \left\langle e^{-\epsilon_1} + e^{-\epsilon_2} \right\rangle_\text{pop}
    \end{split}
\end{gather}
Where $C_1$ and $C_2$ represent the concentrations of the two antigens. As a result of these definitions a cell will survive selection if it is able to bind at least one of the two antigens with good affinity.\\
The combination of these two effects, namely the fact that beneficial mutation will often affect only one of the two components, and selection will allow for survival of cells that have high affinity for at least one of the two antigens, makes so that evolution will split the population in two subsets: cells with high affinity for one antigen or cells with high affinity for the other. To visualize this we simulated the evolution of the system at antigen concentration $C_1 = C_2 = 5$. We performed 500 different simulations. In \cref{fig:mags} we plot the evolution of the average binding energy distribution $P_t(\epsilon_1, \epsilon_2)$ of the population, and the two marginal distributions $P_t(\epsilon_1)$, $P_t(\epsilon_2)$ at four different times $t=0, 10, 50, 100$. The marginal distributions show how the population divides in these two subsets. These subsets evolve independently, each one improving its affinity for their ``target'' antigen over time, while the binding energy for the other antigen gradually decreases due to deleterious mutations not being selected against. The evolution of each of these subsets is therefore well captured by the single-antigen framework along the corresponding axis.\\
We point out that this split in the population is a consequence of the two assumptions we made, on the way mutations and selection operate on the two energy component. Evolution can proceed differently if for example the effect of mutations on the two components is strongly correlated, or if selection requires cells to have high affinity for all of the antigens at the same time.\footnote{The latter case seems less likely given the presence of low-affinity responders in immunization experiments \cite{eyer2020quantitative,kuraoka2016complex}.} If high affinity for both antigens is required then the population will not split in two subsets, but rather evolve along the diagonal of the affinity space $\epsilon_1 = \epsilon_2$. One could therefore simply study the dynamics along a single component, for which the single antigen case would still offer a useful framework.\\

\subsection{Model limitations and perspectives}

Compared to real AM, our model is obviously simplified in many aspects, for example we do not impose an affinity ceiling and beneficial mutations can accumulate indefinitely. However we believe this approximation to be reasonable since for the bottleneck scenario we consider low-affinity cells that could potentially undergo many affinity-improving mutations. Moreover we consider Ag concentration to be constant, while in reality Ag is subject to natural decay and consumption by B-cells. We believe this approximation to be acceptable when studying bottleneck survival, that in most cases is resolved in few evolution rounds. In our analysis we focused on the evolution of the population of B-cells inside a single GC. In the course of a response however many GCs form inside the body.\footnote{While their number is not known with accuracy, it could range from many tens to few hundreds since spleen sections revealed around 20-50 GCs in mice  \cite{wittenbrink2011there}.} By averaging over the distribution of energies in the initial population our theory indeed quantifies the average bottleneck survival probability of GCs in this ensemble, in the hypothesis that the affinity of cells each initial population is independent. If instead affinities are correlated or if cells are able to migrate between GCs then the average would be harder to compute. Unfortunately the lack of precise experimental quantification of these processes forbids any meaningful modeling so far.\\
Finally, our model does not account for the fact that GC selection might be highly permissive. Indeed, low affinity cells have been shown to be able to reside in GCs for a extendend periods of time, and permissiveness might especially characterize maturation against complex antigens \cite{finney2018germinal, molari2020elife}. This effect could be accounted for in our model by adding a small affinity-independent probability of surviving selection. This might have a beneficial impact on maturation, since it could allow low-affinity cells to reside longer in GCs and it might provide them a higher chance of developing beneficial mutations.

\vskip .5cm
\noindent{\bf Acknowledgments:} We are deeply grateful to Jean Baudry, Arup Chakraborty and Klaus Eyer for many  useful discussions and interactions.

\appendix
\section{Model parameters choice}
\label{app:parameters}

\begin{table*}[t]
\centering
\begin{tabular}{r|l|p{10cm}}
  \textbf{parameter}& \textbf{value} & \textbf{description} \\
  \hline
  $\mu_i, \, \sigma_i$& 4, 1.5 &mean and variance of the population initial binding energy distribution \\
  $N_i$& 2500 &initial population size \\
  $N_{\max}$& 2500 & maximum carrying capacity\\
  $p_\mathrm{sil}, \, p_\mathrm{let}, \, p_\mathrm{aa}$& 0.75, 0.15, 0.1 &  probabilities of silent, lethal, affinity-affecting mutations\\
  $\mu_\mathrm{M}, \, \sigma_\mathrm{M}$& 0.3, 0.3& mean and variance of distribution of affinity-affecting mutations\\
  $\epsilon_\text{Ag}$& 0 & Ag-binding selection threshold energy\\
  $C$& see figure captions & Ag concentration \\
  $p_\mathrm{diff}$& 0.1 & differentiation probability
\end{tabular}
\caption{standard values of model parameters. Unless otherwise specified these are the values used in simulations. The choice of their value is discussed in \cref{app:parameters}}.
\label{tab:model_par}
\end{table*}

The values of the parameters are reported in \cref{tab:model_par}, and were chosen based on existing literature.

Mature GCs have a B-cells population consisting of a few thousands cell \cite{jacob1993situ, mcheyzer1993antigen, tas2016visualizing}. We therefore set the initial and maximum size of the population equal to $N_i = N_{\max} = 2500$. This is in agreement with \cite{eisen2014affinity} which reports around 3000 cells per GC. However we stress that GCs are heterogeneous in size \cite{wittenbrink2011there}.
Similarly to \cite{wang2015manipulating, molari2020elife} we consider the duration of a turn of evolution to be $T_\mathrm{round} \sim 12$h, which is consistent with timing of cell migration \cite{victora2010germinal, mesin2016germinal}. Time in our model will be rescaled by this standard quantity, so that the variable $t$ has no dimension. Similarly, also the binding energy $\epsilon$ is dimensionless, expressed in standard units of $k_B T$. For simplicity following experiments that indicate a cell-cycle time of 12h or longer \cite{allen2007imaging} we consider a single division per round. Other experiments indicate an average of two cell division in the DZ \cite{gitlin2014clonal}. We point out that, at the expense of simplicity, our theory can also be extended to account for an higher number of cell divisions.\\
In \cite{wang2017optimal,zhang2010optimality} mutations occur at a rate of 0.5 per sequence per division, and are silent, lethal or affinity affecting with probabilities of respectively $0.5, 0.3, 0.2$. This fixes our effective mutation probabilities to $p_\mathrm{sil} = 0.75, p_\mathrm{let} = 0.15, p_\mathrm{aa} = 0.1$. To reproduce the fact that most of the mutations are deleterious we pick for simplicity $\mu_\mathrm{M} = \sigma_\mathrm{M}$. This fixes the amount of beneficial mutations to $\sim 16\%$, which is somewhat higher but still compatible with other models \cite{zhang2010optimality, wang2015manipulating, molari2020elife} in which this fraction is set to $5\%$. Moreover we set $\mu_\mathrm{M} = 0.3$ so as to set the mean effect of beneficial mutations to $\avg{\Delta \epsilon}_\mathrm{beneficial} \sim -0.15$. This value is slightly smaller than $\avg{\Delta \epsilon}_\mathrm{beneficial} \sim -0.53$ used in \cite{molari2020elife}, but this is compensated by the higher rate of beneficial mutations in our model. As described in the main text, the binding energy distribution of the initial population is set to a Gaussian with standard deviation $\sigma_i=1.5$, which is compatible with experimental data \cite{molari2020elife}. Since evolution of the population is invariant for shifts of the energy space we set $\epsilon_\mathrm{Ag} = 0$. Under this choice of gauge the zero in the energy space is the threshold energy for Ag-binding selection. Moreover we pick $\mu_i = 4$ so that the difference $\mu_i - \epsilon_\mathrm{Ag} - \log C \sim 2 \, \sigma_i$ for the values of Ag concentrations considered in this work ($C \sim 5$) and on average only around $2\%$ of cells from the initial population meet this threshold. For simplicity we independently extract the energy of each cell in the initial population from this initial distribution. By doing so we might overestimate the diversity of the initial population. In fact, experiments probing the clonal composition of GCs estimated that early GCs contain around 50 to 200 different clonal families \cite{tas2016visualizing}. A more realistic initiation of our GCs would require us to extract the energies of around a hundred different founder cells, and let them duplicate without mutating up to to the full GC capacity. This would generate a less homogeneous initial population than the one we consider in our simulation, but would not otherwise strongly impact our results.
Lastly, as in \cite{molari2020elife, wang2015manipulating}, the probability of differentiation is set to $p_\mathrm{diff} = 0.1$.

\section{Estimation of $\bar \epsilon$ evolution}
\label{app:eb}

Including the effect of competition in our evaluation of the population survival probability requires us to estimate the evolution of $\bar \epsilon$, defined in \cref{eq:psurv_t}. To obtain a tractable approximation, we consider the deterministic limit of big population size. In this limit the population binding energy can be approximated with a continuous distribution, and the state of the system is completely determined by the density function $\rho_t(\epsilon)$. This function represents the density of cells having energy $\epsilon$ at evolution round $t$, so that its integral is equal to the size of the population, and its normalized version is the population binding energy distribution. Evolution is expressed in terms of operators acting on this function. In particular:
\begin{enumerate}
	\item Cell duplication corresponds simply to doubling in size:
		\begin{equation}
			\mathbf{A}[\rho](\epsilon) = 2 \times \rho(\epsilon)
		\end{equation}
	\item Mutations are represented as the convolution with the mutation kernel $K (\Delta \epsilon)$ defined in \cref{eq:mut_ker}. Notice that the kernel $K$ is not normalized, to account for the contribution of lethal mutations. It acts on the distribution as:
		\begin{equation}
			\mathbf{M}[\rho](\epsilon) = \int d\Delta\epsilon \, \rho(\epsilon - \Delta \epsilon) \, K (\Delta \epsilon)
		\end{equation}
	\item Ag-binding selection is implemented by in the product of the population function with the survival probability \cref{eq:psurv_ag}:
		\begin{equation}
			\mathbf{S}_\text{Ag}[\rho](\epsilon) = P_\text{Ag}(\epsilon) \, \rho (\epsilon)
		\end{equation}
	\item Similarly, T-cell help selection is given by the product with the survival probability \cref{eq:psurv_t}:
		\begin{equation}\label{eq:tsel_op}
			\mathbf{S}_\text{T}[\rho](\epsilon) = P_\text{T}(\epsilon, \bar\epsilon) \, \rho (\epsilon), \quad \mathrm{with} \; e^{-\bar\epsilon} = \frac{1}{N}\int d\epsilon \, e^{-\epsilon} \, \rho(\epsilon)
		\end{equation}
		Where $N = \int d\epsilon \rho(\epsilon)$ is the current population size.
	\item Differentiation consists simply in a product involving the differentiation probability:
		\begin{equation}
			\mathbf{D}[\rho](\epsilon) = (1 - p_\mathrm{diff}) \, \rho(\epsilon)
		\end{equation}
	\item Finally, the carrying capacity constraint is implemented again by a product and is operated only if the size of the population exceeds the maximum limit:
		\begin{equation}
			\mathbf{C}[\rho](\epsilon) = \min \{1, N_{\max} / N\} \,\rho(\epsilon)
		\end{equation}
		Where again $N = \int d\epsilon \rho(\epsilon)$ is the current population size.
\end{enumerate}
From these definitions the evolution of the population density function $\rho_t(\epsilon)$ can be expressed as:
\begin{equation}
    \rho_{t+1} = \mathbf{C \, D \, S_\mathrm{T} \, S_\mathrm{Ag} \, M \, A} \; \rho_t
\end{equation}
Combining with the definition for $\bar\epsilon$ \cref{eq:tsel_op} provides a way for us to estimate the average evolution of $\bar\epsilon_t$.\footnote{Notice that from the order of the operators in the evolution round it follows that $\epsilon_t$ must be evaluated using \cref{eq:tsel_op} not directly on $\rho_t$ but rather on $\mathbf{S_\mathrm{Ag} \, M \, A} \; \rho_t$}

\section{Derivation and numerical evaluation of recursion equation}
\label{app:rec_explanation}

The core of our theory rests on recursion \cref{eq:dt_recursive}, which allows one to evaluate the probability of extinction of the progeny of a cell from the extinction probabilities of its daughter cells. In this appendix we present an intuitive derivation of this equation, and explain how it was numerically evaluated.\\

As described in the main text, \cref{eq:dt_recursive} operates a recursion on the function $d_t(\epsilon)$, defined as the probability that all the progeny of a given progenitor cell with affinity $\epsilon$ will go extinct by evolution round $t$. The recursion consists in expressing this probability as a function of the progeny extinction probability for the two daughter cells of the progenitor in the remaining $t-1$ rounds. These daughter cells might develop a mutation which introduces a change in binding energy of magnitude $\Delta\epsilon$, and the recursion will therefore contain the term $d_{t-1}(\epsilon+\Delta\epsilon)$. The derivation proceeds in the following manner, schematized in \cref{eq_app:recursion_expl}. The probability that all of the progeny of the progenitor cell with energy $\epsilon$ goes extinct by evolution round $t$ is equal to the probability that both of its daughter cell progenies will go extinct before this round. These probabilities can be expressed as one minus their complement: the probability that extinction does not occur by round $t$. In turn this is equal to the probability that the daughter cell will mutate, survive selection of the first round, and some of its progeny will also survive the remaining $t-1$ rounds. Expressing this latter survival probability in terms of its complement completes the recursive relation.\\
Notice that while the recursion is back in time ($t \to t-1$), the term $d_{t-1}$ refers to the survival probability of the daughter cell, and in this sense the recursion is forward in time. In fact $t$ refers in this case to the number of turns left to survive, and decreases of one unit at every jump to the next generation.\\
\begin{widetext}
\begin{equation}
\label{eq_app:recursion_expl}
    \begin{split}
        d_t(\epsilon) &= P(\text{progeny extinct by round } t)\\
        &= \left[P(\text{daughter cell's progeny extinct by round } t)\right]^2\\
        &= \left[1 - P(\text{daughter cell's progeny not extinct by round } t)\right]^2\\
        &= \left[1 - \int d{\Delta \epsilon} \, P(\text{mutation } \Delta \epsilon) \, P(\text{survive round}) \, P(\text{progeny survives next } t-1 \text{ rounds})\right]^2\\
        &= \left[1 - \int d{\Delta \epsilon} \, P(\text{mutation } \Delta \epsilon) \, P(\text{survive round}) \, (1-d_{t-1}(\epsilon + \Delta \epsilon))\right]^2\\
    \end{split}
\end{equation}
\end{widetext}

Numerical evaluation of \cref{eq:dt_recursive} was performed according to the following scheme. The fact that cells can accumulate mutations makes so that the recursion requires evaluating an integral over all the possible values of the energy. As a consequence in order to evaluate $d_t(\epsilon)$ one needs to know the value of $d_{t-1}(\epsilon)$ at the previous time interval for any value of $\epsilon$. In practice, when evaluating the function at a given final time $T$ one can define a matrix $d[t,i]$, where the first index $t=1,\,\ldots , T$ runs over all possible values of times, and the index $i = 0, \,\ldots,\, I$ operates a discretization of the energy space $\epsilon_0, \ldots, \epsilon_I$, in such a  way that $d[t,i] = d_t(\epsilon_i)$. We indicate with $\delta\epsilon = \epsilon_{i+1} - \epsilon_i$ the value of the discretization step. The first row of the matrix is given simply by the discretized form of \cref{eq:d1} in the main text:
\begin{equation}
    d[1,i] = \left[ 1 - \sum_j \delta \epsilon \, K[j] \, P_S[i + j] \, (1 - p_\text{diff}) \right]^2
\end{equation}
Where $K[i] = K( i \, \delta_\epsilon)$ and $P_S[i] = P_S(\epsilon_i)$ are the discretized forms of the mutation kernel and selection survival probability. From here rows of the matrix can be iteratively populated using the discretized form of \cref{eq:dt_recursive}:
\begin{widetext}
\begin{equation}
    d[t,i] = \left[ 1 - \sum_j \delta\epsilon \, K[j] \, P_S[i + j] \, (1 - p_\mathrm{diff}) \, \left(1 - d[t-1,i+j]) \right)\right]^2
\end{equation}
\end{widetext}
In practice the evaluation of the extinction probability at time $T$ requires of the order of $T$ convolutions between arrays having a size $(\epsilon_I - \epsilon_0) / \delta\epsilon$. Many efficient algorithms exist to perform fast numerical convolution. In our implementation for example, written in Python version 3.9, the evaluation of $d_t$ for $t=150$ and discretizing the binding energy interval $[-50, 50]$ at a discretization step $\delta\epsilon = 0.01$ requires around $0.15$ seconds on a laptop.\\
The more general case of a time-varying survival probability, described in \cref{eq:d_first_BT,eq:d_iterative_BT} in the main text, also presents a similar complexity. In this case one is interested in evaluating the probability $d_{0,T}(\epsilon)$ that a cell having energy $\epsilon$ at time 0 will have its progeny extinct by evolution round $T$. In order to do so one defines a matrix $d[t,i] = d_{t,T}(\epsilon_i)$, with $t=0, \ldots, T-1$. The row of the matrix for $t=T-1$ can be populated from the discretized form of \cref{eq:d_first_BT} in the main text. At this point using the discretized form of \cref{eq:d_iterative_BT} one can populate row $t-1$ from row $t$, until reaching the desired value $t=0$. As in the previous case, this requires of the order of $T$ convolutions of arrays having the size of the discretized binding energy space, and requires a similar computational time.

\bibliography{biblio}

\end{document}